\documentclass[a4paper,11pt]{article}
\pdfoutput=1
\usepackage{jheppub}

\usepackage{graphicx}
\usepackage{dcolumn}
\usepackage{multirow}
\usepackage{bm}
\usepackage[T1]{fontenc}
\usepackage{feynmp-auto}
\usepackage{tabularx}
\usepackage{xcolor}
\usepackage{hyperref}
\usepackage{enumitem}
\usepackage{mathtools}
\usepackage{amsmath}
\usepackage{amssymb}
\usepackage{tikz-feynman}
\usepackage{empheq}
\usepackage{tabu}
\usepackage{tablefootnote}
\usepackage{longtable}
\usepackage{tabularx,booktabs}
\newcolumntype{Y}{>{\centering\arraybackslash}X}
\usepackage{url}

\usepackage{subcaption}

\allowdisplaybreaks

\DeclareUnicodeCharacter{2212}{\ensuremath{-}}
\begin{document}
\preprint{
\vskip-3cm{\baselineskip14pt
    \begin{flushright}
     \normalsize CERN-TH-2025-095\\
     \normalsize CERN-EP-DRAFT-MISC-2025-007
    \end{flushright}} \vskip2cm
}

\title{Time-Dependent Precision Measurement of $B_s^0\to \phi\mu^+\mu^-$ Decay at FCC-$ee$}

\author[a]{Tsz Hong Kwok,} \author[a]{Zachary Polonsky,} \author[a]{Valeriia Lukashenko,} \author[b]{Jason Aebischer,} \author[a]{Ben Kilminster}
\affiliation[a]{Physik-Institut, Universität Zürich
CH-8057 Zürich, Switzerland}
\affiliation[b]{Theoretical Physics Department, CERN, 1211 Geneva 23, Switzerland}

\abstract{
We study the feasibility of measuring time-dependent $C\!P$ violation in the rare flavor-changing neutral current (FCNC) decay $B_s^0 \rightarrow \phi(\rightarrow K^+K^-) \mu^+ \mu^-$ at the FCC-$ee$. In the Standard Model (SM), $C\!P$ violation in this mode arises only at higher orders and is highly suppressed. Extensions of the SM, collectively referred to as New Physics (NP), can introduce additional $C\!P$-violating phases that enhance such effects. The decay $B_s^0 \rightarrow \phi \mu^+ \mu^-$, mediated by the $b \rightarrow s \ell^+ \ell^-$ transition, is therefore a promising probe of NP. The FCC-$ee$, operating as a high-luminosity $Z$-factory, offers an optimal environment for this measurement due to its large event yield, clean conditions, efficient particle identification, and excellent vertex resolution. We perform a Monte Carlo study using \texttt{Pythia} and \texttt{Delphes} with the IDEA detector concept. A relative precision better than $\mathcal{O}(1\%)$ on the branching ratio and $\mathcal{O}(10^{-2})$ on the time-integrated $C\!P$ asymmetry is found to be achievable. We determine the projected sensitivities to the observables $D_f$, $C_f$, and $S_f$, which parameterize time-dependent $C\!P$ violation. In the untagged analysis, a precision of $\mathcal{O}(10^{-1})$ on $D_f$ can be reached. With flavor tagging, sensitivities to $C_f$ and $S_f$ improve to $\mathcal{O}(10^{-2})$. These measurements remain inaccessible to current flavor experiments. Interpreting the results within the Weak Effective Theory provides model-independent constraints on $C\!P$-violating NP. This study demonstrates that FCC-$ee$ enables first-time access to $C\!P$-sensitive observables previously beyond experimental reach.
}


\maketitle

\section{Introduction}
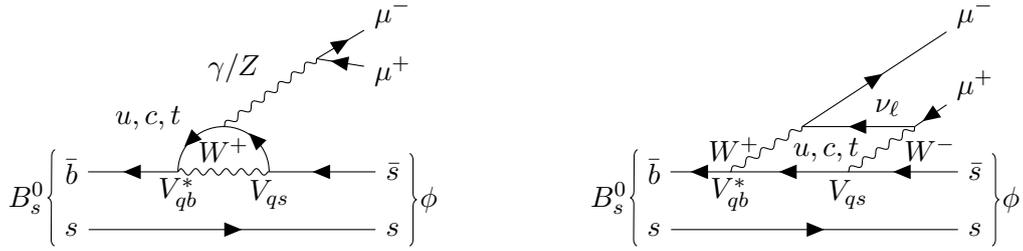
\begin{figure}[h!]
\centering
    \begin{subfigure}[b!]{0.49\textwidth}
        \centering
        \begin{tikzpicture}
        \begin{feynman}
        \vertex (a1) {\(\bar b\)};
        \vertex[right=1.4cm of a1] (a2);
        \vertex[right=1.2cm of a2] (a3);
        \vertex[right=1.4cm of a3] (a4) {\(\bar s\)};
        \vertex[below=2em of a1] (b1) {\(s\)};
        \vertex[below=2em of a4] (b2) {\(s\)};
        \vertex at ($(a1) + (1.4+0.6, 0.6)$) (d);
        \vertex[above=3.5em of a4] (c1) {\(\mu^+\)};
        \vertex[above=2em of c1] (c3) {\(\mu^-\)};
        \vertex (c2) at ($(a4)+(-1,1.5)$);        \vertex (e1) at ($(a2)+(0, -0.3)$) {\(V^{*}_{qb} \)};
        \vertex (e2) at ($(a3)+(0, -0.3)$) {\(V_{qs} \)};
       
        \diagram* {
        (a4) -- [fermion] (a3) -- [boson, edge label'=\(W^+\)] (a2) -- [fermion] (a1),
        (b1) -- [fermion] (b2),
        (c1) -- [fermion] (c2) -- [fermion] (c3),
        (a3) -- [fermion, quarter right] (d) -- [fermion, quarter right, edge label'={\(u,c,t\)}] (a2),
        (d) -- [boson, edge label=\(\gamma / Z\)] (c2),
        };
        \draw [decoration={brace}, decorate] (b1.south west) -- (a1.north west)
        node [pos=0.5, left] {\(B_s^0\)};
        \draw [decoration={brace}, decorate] (a4.north east) -- (b2.south east)
        node [pos=0.5, right] {\(\phi\)};
        \end{feynman}
        \end{tikzpicture}
    \end{subfigure}%
    ~
    \begin{subfigure}[b!]{0.49\textwidth}
        \centering
        \begin{tikzpicture}
        \begin{feynman}
        \vertex (a1) {\(\bar b\)};
        \vertex[right=1.cm of a1] (a2);
        \vertex[right=1.55cm of a2] (a3);
        \vertex[right=1.45cm of a3] (a4) {\(\bar s\)};
        \vertex[below=2em of a1] (b1) {\(s\)};
        \vertex[below=2em of a4] (b2) {\(s\)};
        
        \vertex at ($(a1) + (1.2+0.75, 0.6)$) (d1);
        \vertex at ($(a2) + (1.68+0.75, 0.6)$) (d2);

        \vertex[above=3.em of a4] (c1) {\(\mu^+\)};
        \vertex[above=2.5em of c1] (c3) {\(\mu^-\)};
        
        \vertex (c2) at ($(a4)+(-1,1.5)$);        \vertex (e1) at ($(d1)+(-0.9, -0.3)$) {\(W^+\)};
        \vertex (e2) at ($(d2)+(0.2, -0.3)$) {\(W^-\)};
        \vertex (e1) at ($(a2)+(0, -0.3)$) {\(V^{*}_{qb} \)};
        \vertex (e2) at ($(a3)+(0, -0.3)$) {\(V_{qs} \)};
        
        \diagram* {
        (a4) -- [fermion] (a3) -- [fermion, edge label'={\(\qquad\;\; u,c,t\)}] (a2) -- [fermion] (a1),
        (b1) -- [fermion] (b2),
        (a2) -- [boson] (d1),
        (a3) -- [boson] (d2),
        (d2) -- [fermion, edge label'=\(\qquad\nu_\ell\)] (d1),
        (d1) -- [fermion] (c3),
        (c1) -- [fermion] (d2),
        };
        \draw [decoration={brace}, decorate] (b1.south west) -- (a1.north west)
        node [pos=0.5, left] {\(B_s^0\)};
        \draw [decoration={brace}, decorate] (a4.north east) -- (b2.south east)
        node [pos=0.5, right] {\(\phi\)};
        \end{feynman}
        \end{tikzpicture}
    \end{subfigure}
    \caption{Examples of leading-order $B_s^0\to \phi\mu^+\mu^-$ Feynman diagrams in the SM. \label{fig:SMFD}}
\end{figure}

The study of rare processes plays a crucial role in probing the limits of the Standard Model (SM) and searching for potential New Physics (NP). One of the primary motivations for studying rare decays is their sensitivity to NP. Any deviation from the SM predictions could indicate the presence of new particles or interactions not accounted for in the SM. In this study, we focus mainly on the rare $b\to s\ell^+\ell^-$ decay, which is mediated by a flavor-changing neutral current (FCNC). Such a FCNC decay is forbidden at tree-level in the SM and can only occur via loop corrections. Figure~\ref{fig:SMFD} shows the rare decay $B_s^0\to\phi\mu^+\mu^-$,\footnote{We adopt a notation in which the inclusion of the charge-conjugation mode is understood, throughout this paper.} which has a branching ratio of $\mathcal{O}(10^{-6})$ at leading order in the SM. This process is sensitive to several beyond-the-SM (BSM) scenarios, such as leptoquark or $Z'$ models, which can contribute to the transition at tree level~\cite{Buttazzo:2017ixm, Boucenna:2016qad, Chiang:2017hlj, Kumar:2018kmr, Barbieri:2017tuq}.

Beyond the search for NP, studying $C\!P$ violation is crucial, as it is one of the Sakharov conditions required to explain the matter--antimatter asymmetry through baryogenesis~\cite{Sakharov:1967dj}. The SM introduces $C\!P$ violation through a single complex phase in the Cabibbo-Kobayashi-Maskawa (CKM) matrix, which describes the strength of the quark transitions~\cite{Aleksan:1994if, Kobayashi:1973fv}.\footnote{The $\theta$-angle in QCD is omitted in the discussion.} The amount of $C\!P$ violation predicted by the SM is insufficient to account for the observed dominance of matter over antimatter, suggesting the presence of additional sources of $C\!P$ violation~\cite{Planck:2018vyg, Jarlskog:1985ht}. This discrepancy motivates the search for BSM $C\!P$-violating sources.

\begin{figure}[h!]
\centering
    \begin{subfigure}[b!]{0.49\textwidth}
        \centering
        \begin{tikzpicture}
        \begin{feynman}
        \vertex (a1) ;
        \vertex[right=-1.1cm of a1]  (a2_)  {\( s\)};
        \vertex[right=2em of a2_] (a2_W1);
        \vertex[right=2em of a2_W1] (a2_W2);
        \vertex[below=2em of a2_W1] (b2_W1);
        \vertex[below=2em of a2_W2] (b2_W2);
        \vertex[right=1.4cm of a1] (a2);
        \vertex[right=1.2cm of a2] (a3);
        \vertex[right=1.4cm of a3] (a4) {\( s\)};
        \vertex[below=2em of a1] (b1);
        \vertex[below=2em of a2_] (b2_) {\(\bar b\)};
        \vertex[below=2em of a4] (b2) {\(\bar s\)};
        \vertex at ($(a1) + (1.4+0.6, 0.6)$) (d);
        \vertex[above=3.5em of a4] (c1) {\(\mu^+\)};
        \vertex[above=2em of c1] (c3) {\(\mu^-\)};
        \vertex (c2) at ($(a4)+(-1,1.5)$);        \vertex (e1) at ($(a2)+(0, -0.3)$) {\(V_{qb} \)};
        \vertex (e2) at ($(a3)+(0, -0.3)$) {\(V^{*}_{qs} \)};

        \diagram* {
        (a2_W2) -- [fermion] (a2) -- [boson, edge label=\(W^+\)] (a3) -- [fermion] (a4),
        (a2_W1) -- [boson, edge label=\(W^-\)] (a2_W2),
        (a2_) -- [fermion] (a2_W1),
        (b2) -- [fermion] (b2_W2),
        (b2_W1) -- [boson, edge label'=\(W^+\)] (b2_W2),
        (b2_W1) -- [fermion] (b2_),
        (a2_W1) -- [fermion] (b2_W1),
        (b2_W2) -- [fermion] (a2_W2),
        
        (c1) -- [fermion] (c2) -- [fermion] (c3),
        
        (a2) -- [fermion, quarter left, edge label={\(\bar u,\bar c,\bar t\)}] (d) -- [fermion, quarter left] (a3),
        (d) -- [boson, edge label=\(\gamma / Z\)] (c2),
        };
        \draw [decoration={brace}, decorate] (b2_.south west) -- (a2_.north west)
        node [pos=0.5, left] {\(B_s^0\)};
        \draw [decoration={brace}, decorate] (a4.north east) -- (b2.south east)
        node [pos=0.5, right] {\(\phi\)};
        \end{feynman}
        \end{tikzpicture}
    \end{subfigure}
    \caption{Example of leading order $B_s^0\to \phi\mu^+\mu^-$ Feynman diagram in the SM, showing an oscillation of $B_s^0\to\bar{B}_s^0$, which introduces additional phases and interferes with the example in Figure~\ref{fig:SMFD}.\label{fig:SMFD_osc}}
\end{figure}
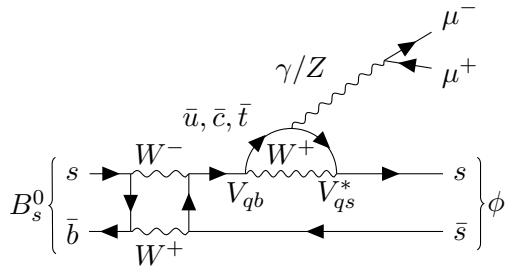

The rare decay $B_s^0\to \phi(\to K^+ K^-)\mu^+\mu^-$, which decays into a $C\!P$ eigenstate, can give rise to $C\!P$ violation in interference between decays with and without neutral meson mixing. For the mixing case, Figure~\ref{fig:SMFD_osc} illustrates an example of an oscillating process, where a $B_s^0$ meson oscillates into a $\bar{B}_s^0$ and subsequently decays to $\phi\mu^+\mu^-$. This behavior produces a characteristic signature in the time-dependent $C\!P$ asymmetry. Measuring the $C\!P$ properties of such suppressed decays is especially informative, as even a small NP contribution to the weak phase can have a significant impact relative to the small SM amplitude. Potential NP effects can be studied using a model-independent Effective Field Theory (EFT) framework, where the imaginary parts of the relevant Wilson Coefficients (WCs) encode possible $C\!P$-violating contributions from short-distance interactions. Most current $C\!P$ violation studies focus on non-leptonic decays, where hadronic uncertainties can obscure potential NP contributions. In comparison, semi-leptonic decays such as flavor-changing neutral current (FCNC) transitions provide cleaner theoretical predictions due to reduced hadronic effects and better-controlled form factors. Measuring $C\!P$ violation in these channels therefore offers a complementary and theoretically robust test of the SM. On the theory side, several studies have examined methods for extracting $C\!P$-violating effects from FCNC processes~\cite{Descotes-Genon:2022qce, Bobeth:2008ij, Fleischer:2024fkm, Fleischer:2022klb, Descotes-Genon:2015hea, Bobeth:2012vn, Ai:2024nmn, Fleischer:2025ucq}. 

In this study, we interpret potential NP effects using the Weak Effective Theory (WET), which describes low-energy interactions after integrating out heavy electroweak-scale particles. Specifically, this study focuses on the WCs $C_7$, $C_9$, and $C_{10}$, analyzing both their real and imaginary components to probe potential sources of $C\!P$ violation. Their definitions are given in Section~\ref{sec:inter}.

To enable time-dependent measurements in the rare $B_s^0 \to \phi\mu^+\mu^-$ decay, a substantial and relatively background-free dataset is required, particularly to facilitate flavor tagging. Collecting such a dataset to enable time-dependent measurements with meaningful precision is, however, beyond the reach of current experimental facilities. This poses significant challenges for existing $B$-physics programs at contemporary colliders. Specifically, the collision energies of SuperKEKB are not optimized for producing $B_s^0$ mesons, resulting in a statistically limited $B_s^0$ dataset at Belle II, which relies on dedicated data-taking runs at the $\Upsilon(5S)$ resonance. On the other hand, while LHCb can produce a larger number of heavy hadrons, measuring time-dependent $C\!P$ violation is challenging due to the high levels of QCD background at the LHC. This limits the achievable flavor tagging power, which refers to the ability to determine whether the decay originated from a $B_s^0$ or $\bar{B}_s^0$, as well as the precision of vertex reconstruction and proper decay time measurement. In contrast, future $Z$-factories like the FCC-$ee$~\cite{FCC:2018byv, FCC:2018evy, Bernardi:2022hny} and CEPC~\cite{CEPCStudyGroup:2023quu, CEPCStudyGroup:2018ghi, CEPCStudyGroup:2018rmc} provide an ideal environment for conducting such measurements. 

Such a challenging precision measurement is made possible in future $Z$-factories by several factors. Firstly, the $Z$-pole run of the FCC-$ee$ is expected to produce $\sim 6\times 10^{12}$ on-shell $Z$ bosons, resulting in up to $\sim 1.9\times 10^{11}$ $B_s^0$ and $\bar{B}_s^0$ mesons, as detailed in Table~\ref{tab:Bnum} by the expected yields at Belle~II~\cite{Belle-II:2018jsg}, the LHCb Upgrade II~\cite{LHCb:2018roe} and the FCC-$ee$. Secondly, due to its leptonic initial state, the FCC-$ee$ provides a significantly cleaner experimental environment compared to a hadronic machine. This not only enhances background suppression more effectively but also improves the particle identification and flavor tagging power~\cite{Blekman:2024wyf}. Additionally, operating at $\sqrt{s}=m_Z$ ensures a high boost of the produced particles, thereby enhancing displaced vertex significance. A good resolution in vertexing is crucial for accurate time-dependent $C\!P$ violation measurements. A more boosted displaced vertex, combined with advanced detector design features such as a lower material budget, consequently improves the precision of $b$-hadron vertex reconstruction. Recent studies exploring flavor physics potential at future $Z$-factories include, but are not limited to, Refs.~\cite{Ho:2022ipo, Zheng:2020ult, Li:2022tov, Liang:2024hox, Li:2022tlo, Aleksan:2021gii, Aleksan:2021fbx, Amhis:2021cfy, Kamenik:2017ghi, Li:2020bvr, Monteil:2021ith, Chrzaszcz:2021nuk, Dam:2018rfz, Qin:2017aju, Li:2018cod, Calibbi:2021pyh, Bordone:2025cde, Allwicher:2025bub}.

\begin{table}
\centering
\begin{tabular}{ccccc}
\hline 
\hline 
   $b$-hadron & Belle~II & LHCb & FCC-$ee$\\
\hline 
$B^0$, $\bar{B}^0$ & $5.3\times 10^{10}$ & $ 6\times 10^{13}$  & $7.2 \times 10^{11}$ \\
$B^\pm$ & $5.6\times 10^{10}$ & $ 6\times 10^{13}$  & $7.2 \times 10^{11}$ \\
$B_s^0$, $\bar{B}_s^0$ & $5.7 \times 10^{8}$ & $ 2\times 10^{13}$  & $1.9\times 10^{11}$ \\
$B_c^\pm$ & -- & $ 4 \times 10^{11}$  & $1.1\times 10^9$ \\
$\Lambda_b^0$, $\bar{\Lambda}_b^0$ & -- & $ 2\times 10^{13}$  & $1.5\times 10^{11}$  \\
\hline
\hline 
\end{tabular}
\caption{Estimated yield of different $b$-hadrons procured at Belle~II, LHCb and the {FCC-$ee$}~\cite{Ho:2022ipo}.} 
\label{tab:Bnum}
\end{table}

Other decay channels can also be used to study the $C\!P$ properties of $b \to s\mu^+\mu^-$ processes~\cite{Fleischer:2022klb}. These include, for example, the decays $B^+ \to K^+ \mu^+ \mu^-$, $B^0 \to K^{*0} \mu^+ \mu^-$, and $B^0 \to K_S^0 \mu^+ \mu^-$. In addition, $C\!P$ violation studies have also been proposed for rare FCNC $b \to s \nu \bar{\nu}$ decays~\cite{Amhis:2023mpj, Descotes-Genon:2022gcp}, including $B^0 \to K^{*0} \nu \bar{\nu}$, $B_s^0 \to \phi \nu \bar{\nu}$, $B^0 \to K_S^0 \nu \bar{\nu}$, and $\Lambda_b^0 \to \Lambda \nu \bar{\nu}$. In principle, the time-dependent measurement techniques explored in this study can also be applied to these channels, accounting for the interference between decays with and without neutral meson mixing. However, in this work we focus on the decay $B_s^0 \to \phi(\to K^+ K^-)\mu^+\mu^-$ for several reasons. First, we aim to demonstrate the full potential of the FCC-$ee$ in comparison to other experiments. As shown in Table~\ref{tab:Bnum} and discussed above, the increase in $B_s^0$ meson yields at FCC-$ee$ compared to Belle~II is expected to be greater than for other $b$-hadron species. Second, the $\phi(\to K^+K^-)\mu^+\mu^-$ final state is a linear combination of $C\!P$ eigenstates, which is not the case for $B^+ \to K^+ \mu^+ \mu^-$ or $B^0 \to K^{*0}(\to K^+ \pi^-)\mu^+ \mu^-$ decays. While $B^0 \to K_S^0 \mu^+ \mu^-$ also leads to a $C\!P$ eigenstate, its more complex topology, featuring a third detached vertex, makes the analysis more complicated. Therefore, we focus first on a simpler topology. To demonstrate the potential of FCC-$ee$ for time-dependent $C\!P$ violation measurements, we select $B_s^0 \to \phi\mu^+\mu^-$ as the flagship analysis channel. While $b \to s\nu\bar{\nu}$ decays are theoretically cleaner, their experimental reconstruction is significantly more difficult due to the presence of two undetected neutrinos, which degrades vertex resolution. In contrast, the $\phi\mu^+\mu^-$ final state provides visible decay products, enabling precise vertexing and making it more suitable for time-dependent analysis at FCC-$ee$.

This paper is structured as follows. Section~\ref{sec:SimSel} discusses the strategies used to simulate signal and background samples, as well as the selection criteria employed to isolate signal-like events. Section~\ref{sec:results} presents the key experimental observables along with their expected uncertainties, obtained from both untagged and tagged measurements. Section~\ref{sec:inter} interprets these results within the framework of the WET.

\section{Simulation and Event Selection\label{sec:SimSel}}
In this study, we use \texttt{Pythia 8}~\cite{Bierlich:2022pfr} to simulate both signal and background samples for $e^+e^- \to Z$ production at the $Z$-pole. The signal samples are generated from $ Z \to b\bar{b}$ at the $Z$-pole, with the exclusive decay of $B_s^0\to \phi(\to K^+K^-)\mu^+\mu^-$. For the background, we simulate the SM processes of $Z\to b\bar{b}$, where the signal mode is excluded in this sample, and $Z\to c\bar{c}$ inclusively, as they were shown to be the dominant $Z\to q\bar{q}$ background types in similar $B$-physics studies at future $Z$-factories \cite{Li:2022tov}. The detector effects are included using \texttt{Delphes 3}~\cite{deFavereau:2013fsa} with the IDEA detector concept~\cite{IDEAStudyGroup:2025gbt, Antonello:2020tzq, Ilg:2023pmt}. In our study, the vertex detector plays the most crucial role due to its impact on vertex resolution and decay time measurements. The IDEA vertex detector employs monolithic active pixel sensors (MAPS), surrounding a beam pipe with an outer radius of 11.7~mm. The detector features three inner barrel layers, each assumed to have a single-hit resolution of 3~$\mu$m, with the first layer positioned at a radius of 1.37~cm. These are followed by two outer barrel layers, which have an assumed single-hit resolution of 7~$\mu$m. Additionally, the forward and backward regions are each covered by three disk layers, also with an assumed single-hit resolution of 7~$\mu$m.

To better understand the physics of the inclusive backgrounds, they can be classified into three main categories: resonance, cascade and combinatoric mis-identified (mis-ID) backgrounds. Resonance backgrounds represent the physical decays where the $B_s^0$ decays to only $K^+K^-\mu^+\mu^-$ final states, but with $\mu^+\mu^-$ originating from resonances like $\phi$, $J/\psi$ and $\psi(2S)$. An example of such resonance backgrounds is $B_s^0\to \phi(\to K^+K^-) J/\psi(\to\mu^+\mu^-)$. Cascade and combinatoric backgrounds are defined similar to those in Ref.~\cite{Ho:2022ipo}. Specifically, cascade background refers to cases where the final states $K^+ K^- \mu^+ \mu^-$ are produced via intermediate particles decaying from the same $b$-hadron, but which are not classified as resonance backgrounds. One example is the decay $B_s^0\to D_s^- \mu^+\nu_\mu$ followed by $D_s^-\to\phi(\to K^+ K^-)\mu^- \bar{\nu}_\mu$. Combinatoric background represents cases where the final states $K^+ K^- \mu^+ \mu^-$ are not produced from the same hadron. Instead, they are produced directly from fragmentation, or decaying from products of fragmentation. Lastly, mis-ID background arises when individual particles in the event, such as $\pi^\pm$, are incorrectly identified as target particles like $\mu^\pm$, causing the event to mimic the signal final state and pass the selection criteria. However, we neglect this background in our study, given the low misidentification rates expected at future $Z$-factories, as discussed in Appendix~\ref{app:misID}.

We define the following selection criteria to reconstruct and identify signal-like candidates, based on the characteristics of the signal and the backgrounds discussed above:
\begin{itemize}
    \item \textbf{Final States (FS)}: An event should have at least one pair of oppositely charged $K^+ K^-$ tracks and at least one pair of oppositely charged $\mu^+ \mu^-$ tracks. All candidate tracks need to have $p_T>0.1$~GeV. Among these $K^+ K^- \mu^+ \mu^-$ candidates, we require at least one combination to be traveling in the same direction. Specifically, we require that the pairwise dot products of their 3-momenta are positive: $\vec{p}_i\cdot\vec{p}_j\geq 0$ for $i,j\in\{K^+, K^-, \mu^+, \mu^-\}$. Next, we use the tracks of the $K^+ K^- \mu^+ \mu^-$ candidate to perform a vertex fit. If more than one combination satisfy the directional requirement, we select the combination that gives the lowest vertex $\chi^2$ value.
    
    \item \textbf{Vertex fit $\chi^2$}: The selected $K^+ K^- \mu^+ \mu^-$ candidate must have a vertex fit with $\chi^2<15$. The fitted vertex is denoted as $\vec{s}$ and corresponds to the $B_s^0$ decay vertex.
    
    \item \textbf{Momentum $|\vec{p}|$}: Each of the selected $K^+,\ K^-,\ \mu^+$ and $\mu^-$ candidate must have momentum larger than 2~GeV.
    
    \item \textbf{Reconstructed $\phi$ mass $m_{\phi}$}: We first define the reconstructed $\phi$ mass as the invariant mass of the $K^+ K^-$ system: $m_\phi\equiv m_{K^+ K^-}$. We then require the invariant mass to fall within the range $1$~GeV $<m_\phi< 1.04$~GeV.
    
    \item \textbf{Squared dimuon invariant mass $q^2$}: We require that $q^2$ is not in the $\phi$, $J/\psi$ and $\psi(2S)$ resonance regions. Namely, we select ranges $q^2\not\in [1, 1.08]~\text{GeV}^2$, $q^2\not\in [9, 10.2]~\text{GeV}^2$ and $q^2\not\in [13, 14.4]~\text{GeV}^2$. 

    \item \textbf{Reconstructed $B_s^0$ mass $m_{B_s^0}$}: We define the reconstructed $B_s^0$ mass as the invariant mass of the $K^+ K^- \mu^+ \mu^-$ system: $m_{B_s^0}\equiv m_{K^+ K^-\mu^+ \mu^-}$. We then require $5.35$~GeV $<m_{B_s^0}< 5.39$~GeV.

\end{itemize} 

\begin{figure*}[h!]
    \centering
    \begin{subfigure}[t]{0.5\textwidth}
        \centering
        \includegraphics[width=0.95\textwidth]{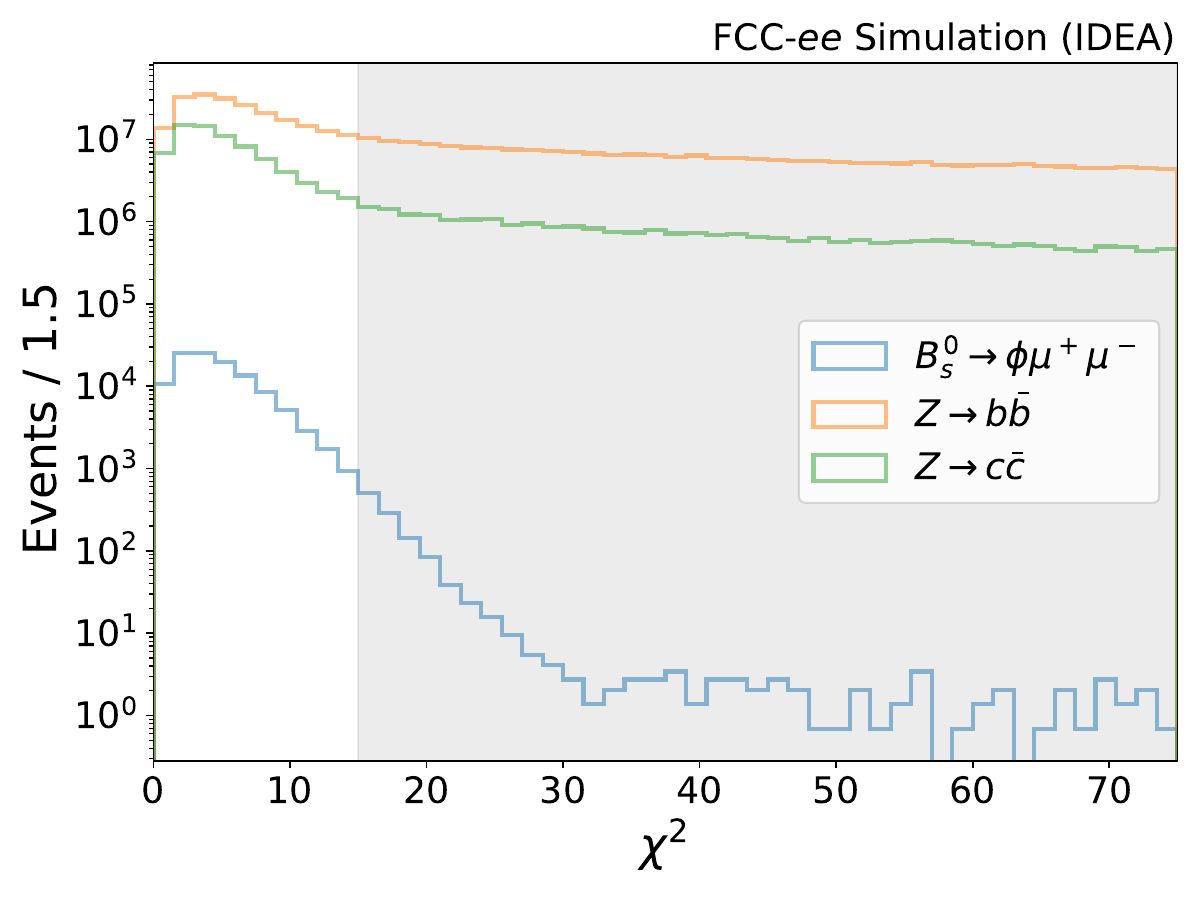}
    \end{subfigure}%
    ~ 
    \begin{subfigure}[t]{0.5\textwidth}
        \centering
        \includegraphics[width=0.95\textwidth]{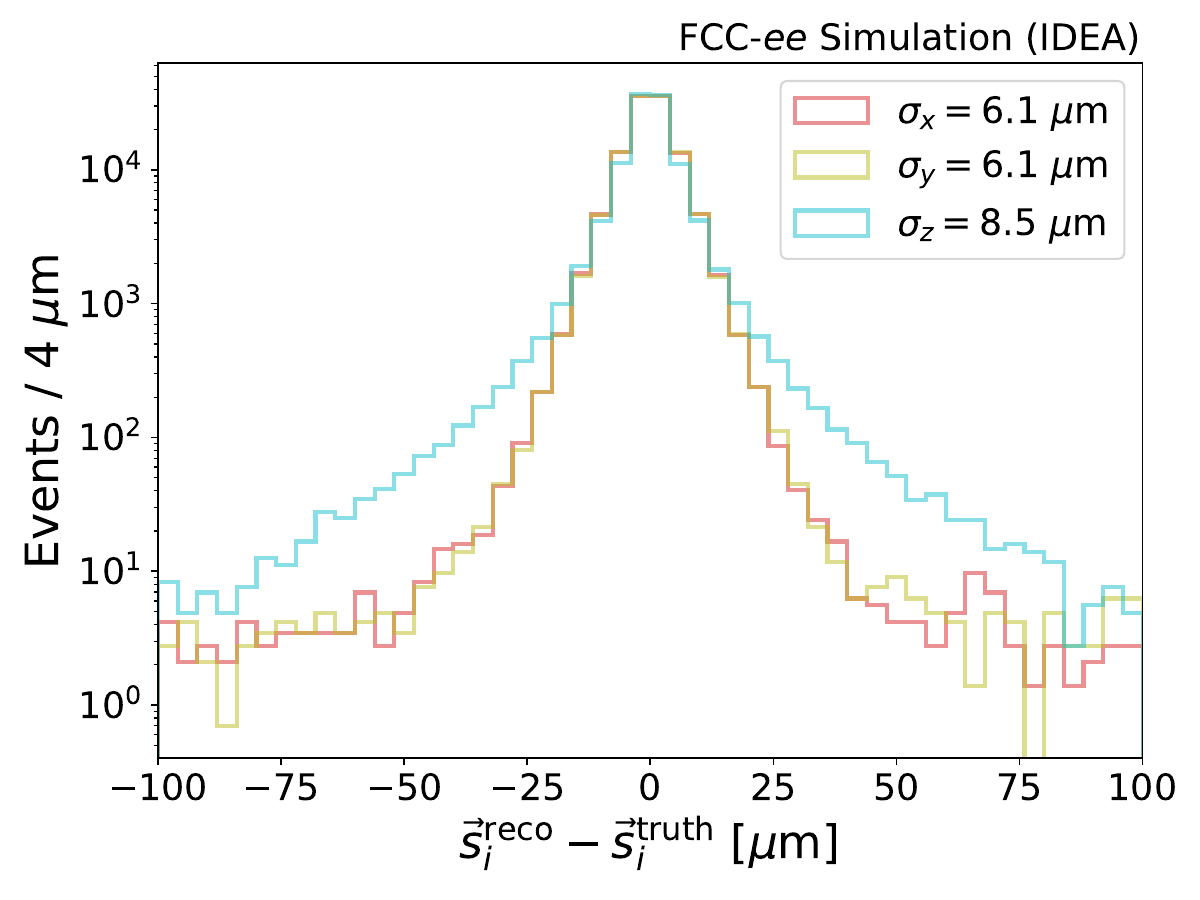}
    \end{subfigure}%

    \caption{\textbf{Left}: Vertex fit $\chi^2$ distribution from the $K^+ K^- \mu^+ \mu^-$ tracks. \textbf{Right}: The residual of the reconstructed secondary ($B_s^0$ decay) vertex vector $\vec{s}$ after the $\chi^2$ selection requirement. Different colors represent the residual projected to the $x,\ y$ and $z$ axes, respectively.}
    \label{fig:vertex}
\end{figure*}

\begin{figure*}[h!]
    \centering
    \begin{subfigure}[t]{0.5\textwidth}
        \centering
        \includegraphics[width=0.95\textwidth]{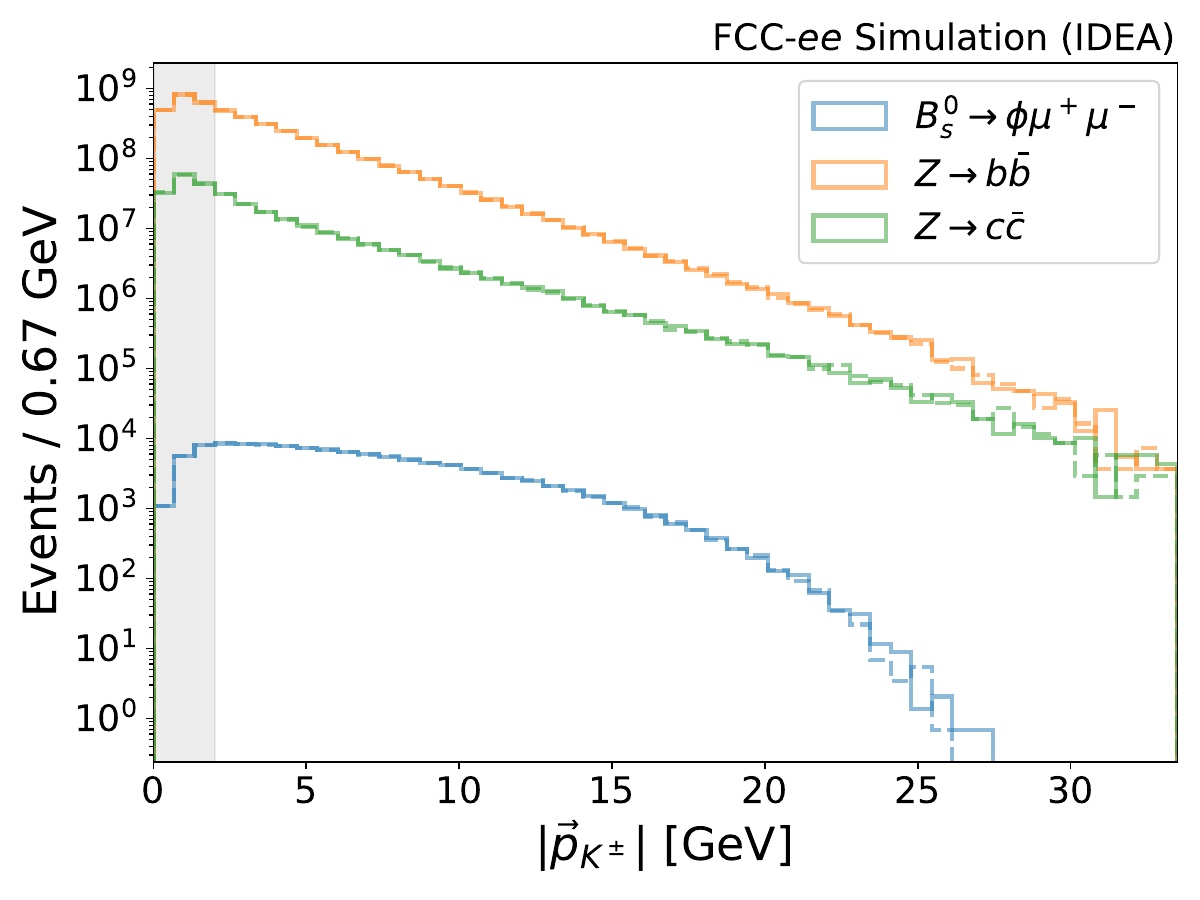}
    \end{subfigure}%
    ~ 
    \begin{subfigure}[t]{0.5\textwidth}
        \centering
        \includegraphics[width=0.95\textwidth]{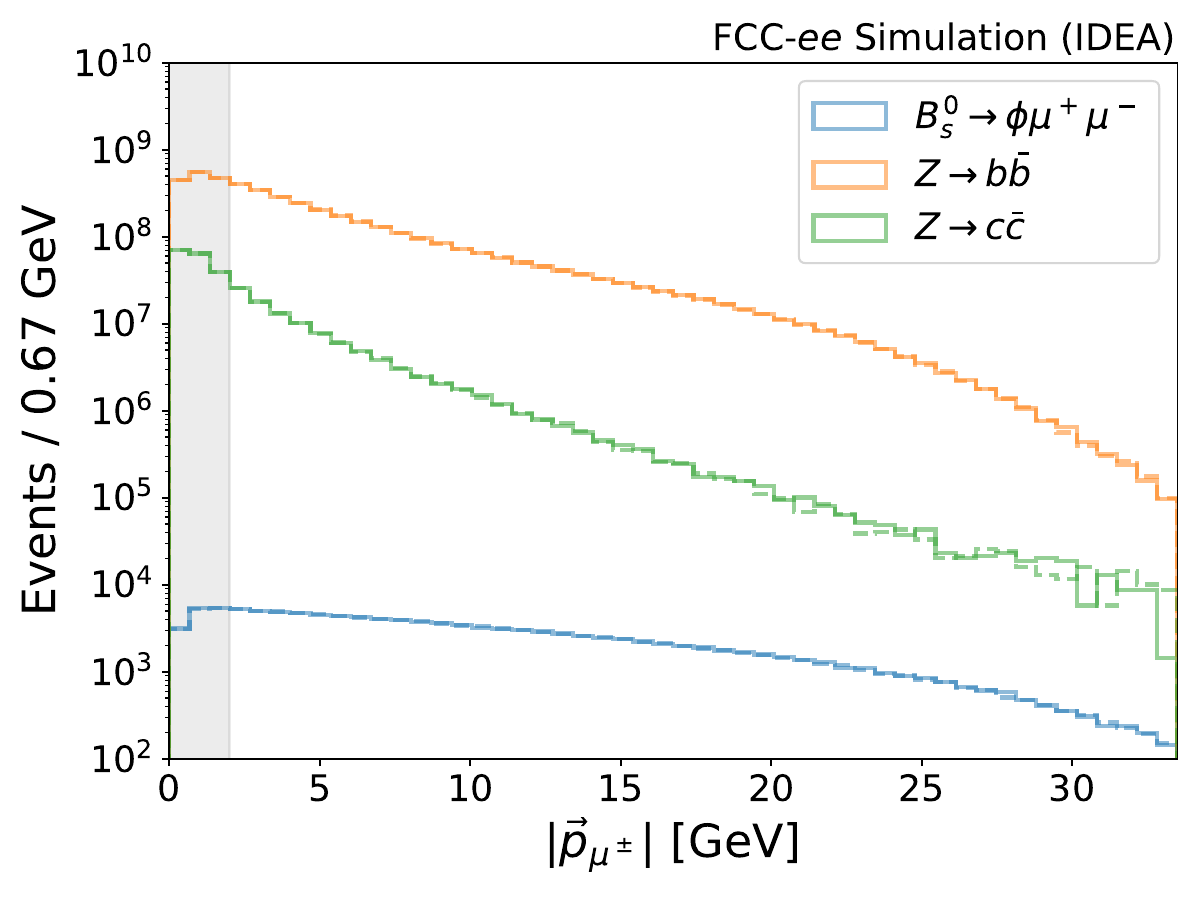}
    \end{subfigure}%

    \caption{The distributions of the momentum of the $K^+,\ K^-,\ \mu^+$ and $\mu^-$ candidates. Solid lines represent $K^+$ or $\mu^+$, and dashed lines represent $K^-$ or $\mu^-$. }
    \label{fig:mon}
\end{figure*}

\begin{figure*}[h!]
    \centering
    \begin{subfigure}[t]{0.5\textwidth}
        \centering
        \includegraphics[width=0.95\textwidth]{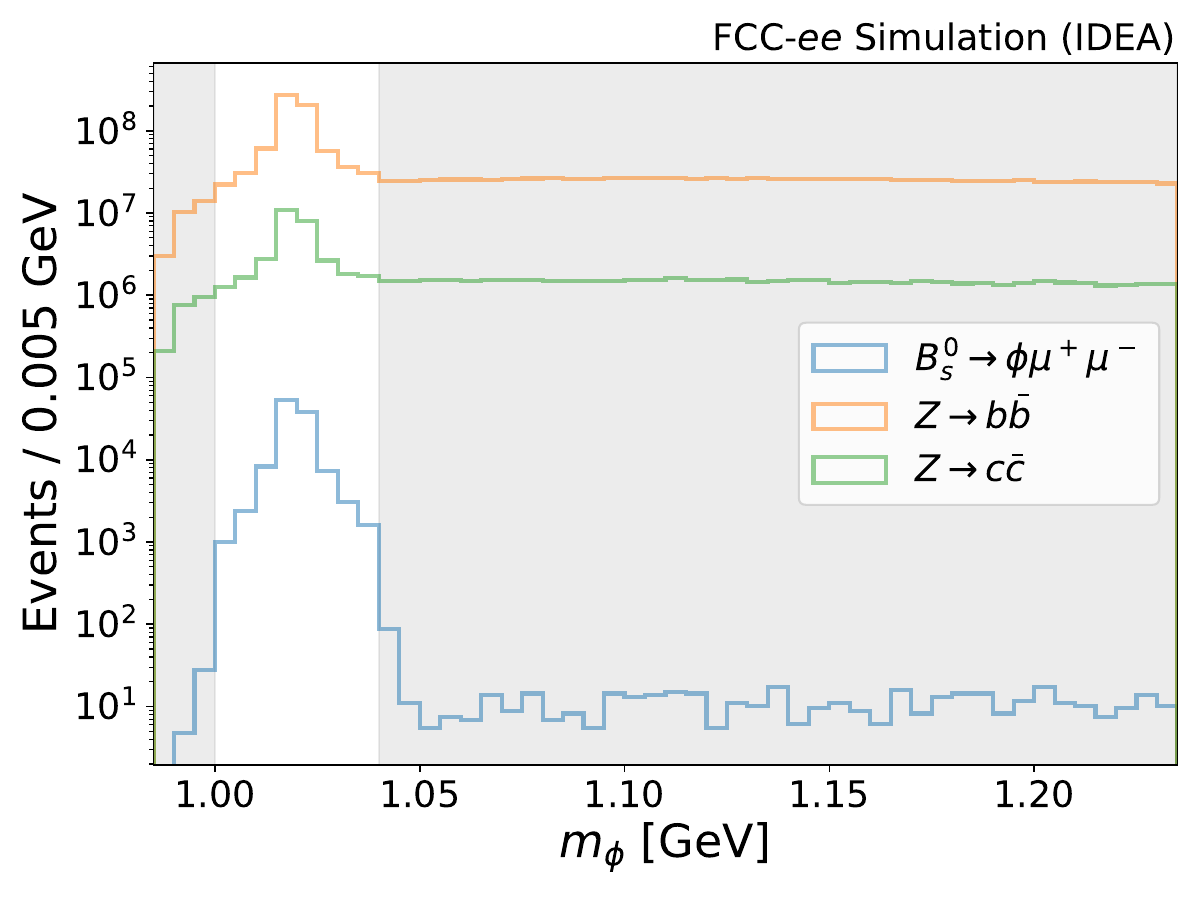}
    \end{subfigure}%
    ~
    \begin{subfigure}[t]{0.5\textwidth}
        \centering
        \includegraphics[width=0.95\textwidth]{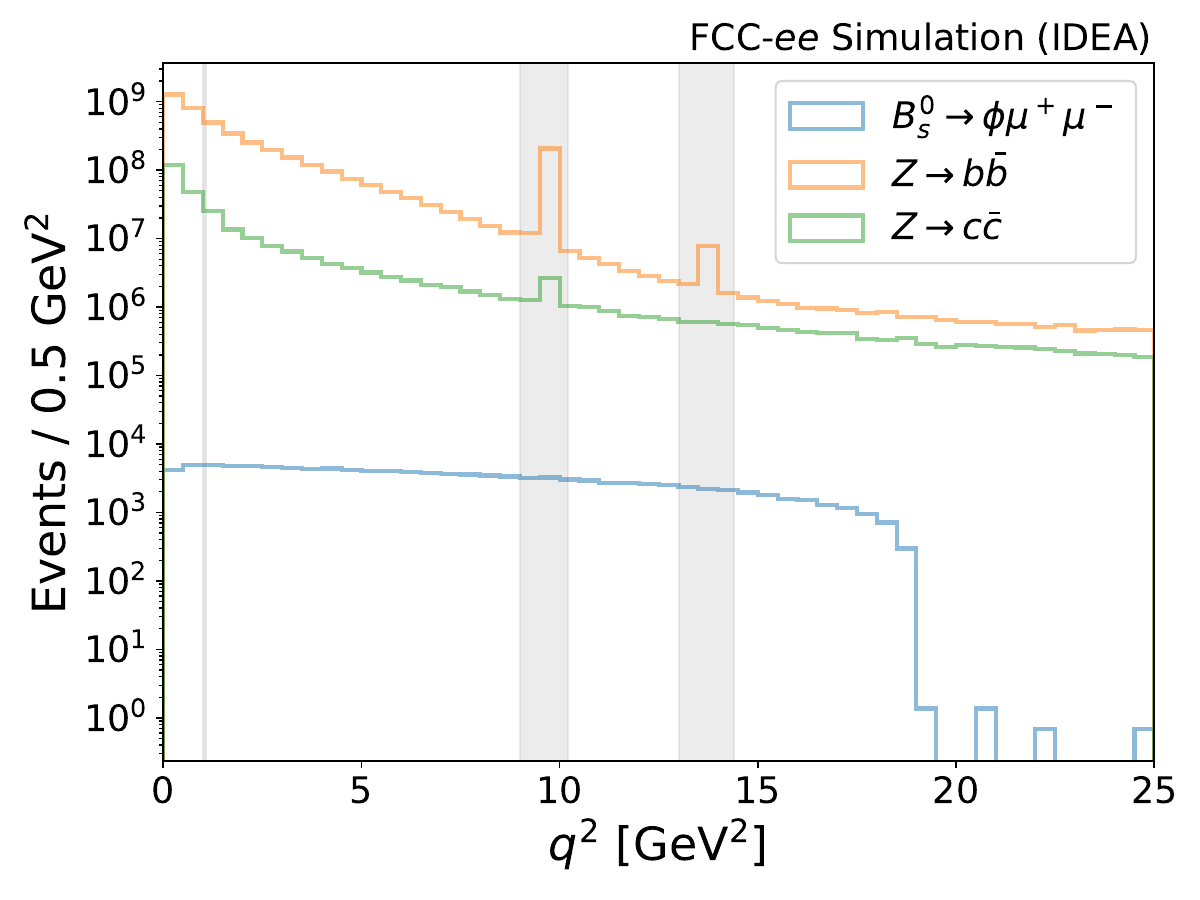}
    \end{subfigure}%

    \caption{\textbf{Left}: The reconstructed $\phi$ mass distributions. \textbf{Right}: The squared dimuon invariant mass distributions.}
    \label{fig:reson}
\end{figure*}

The numbers of events at each selection stage is shown in Table~\ref{tab:cutFlow}. Distributions of the corresponding observables, after the final states selection stage, are also shown in Figures~\ref{fig:vertex},~\ref{fig:mon} and~\ref{fig:reson}. The shaded region in the figures are excluded by the selection.

\begin{table}[h!]
    \centering
    \begin{tabular}{cccc}
        \hline  
        \hline
        Channel & $B_s^0 \to \phi \mu^+\mu^-$ & $Z\to b\bar{b}$ & $Z\to c\bar{c}$ \\
        \hline  
        Events at FCC-$ee$ & $7.82\times 10^{4}$ & $9.07\times 10^{11}$ & $7.22\times 10^{11}$\\ 
        \hline
        $N_{\text{FS}}$ &   $7.30\times 10^4$ & $4.34\times 10^9$ & $2.82\times 10^8$\\
        $N_{\chi^2}$ &   $7.19\times 10^4$ & $2.15\times 10^8$ & $7.25\times 10^7$\\
        $N_{|\vec{p}|}$ & $4.61\times 10^4$ & $5.98\times 10^7$ & $2.25\times 10^6$\\
        $N_{m_{\phi}}$ & $4.59\times 10^4$ & $3.21\times 10^7$ & $3.64\times 10^5$\\
        $N_{q^2}$ &      $3.97\times 10^4$ & $1.24\times 10^7$ & $3.13\times 10^5$\\
        $N_{m_{B_s^0}}$ &  $3.93\times 10^4$ & $1.39\times 10^3$ & $2.13\times 10^2$\\ 
        \hline
        \hline
        
    \end{tabular}
    \caption{Event yields at FCC-$ee$ after successive application of selection criteria. The branching ratio Br($B_s^0\to \phi \mu^+ \mu^-$) uses the world averaged value from HFLAV~\cite{HeavyFlavorAveragingGroupHFLAV:2024ctg}. Note that the branching ratio Br($\phi\to K^+ K^-$) is included in the estimated signal yield.}
    \label{tab:cutFlow}
\end{table}

Figure~\ref{fig:vertex} shows that the signal events are mostly concentrated in the low $\chi^2$ region. On the other hand, the $Z\to b\bar{b}$ (excluding signal) and $Z\to c\bar{c}$ backgrounds exhibit a flat, long tail extending to the large $\chi^2$ region. This feature can be explained by the cascade and combinatoric types of backgrounds introduced above. For cascade backgrounds, the particles may undergo a cascade decay chain involving displaced intermediate particles (such as $D$ mesons), leading to spatially separated final state tracks. Additionally, in combinatoric background, one or more of the final-state particles may originate from fragmentation, resulting in tracks that are spatially displaced from the rest of the final-state tracks. Figure~\ref{fig:vertex} also shows the resolution of the reconstructed secondary vertex in signal samples in the $x, y$ and $z$ directions, which suggests that the resolution of the vertexing is $\mathcal{O}(10~\mu$m$)$. Note that the resolutions in the $x$ and $y$ directions are similar, while that in the $z$ direction is larger. This is primarily because the uncertainty of the vertex is larger along the particle's longitudinal direction, while $b\bar{b}$ is produced more often along the beam direction at $Z$-factories. Furthermore, the resolution in the $z$-direction is $\eta$-dependent, leading to larger uncertainties in the forward region.

The momentum distributions of the four final state particles are shown in Figure~\ref{fig:mon}. The backgrounds are more concentrated in the low momentum region compared to the signal, making momentum-based selection criteria effective for background suppression. Additionally, as demonstrated in Appendix~\ref{app:misID}, particle identification performance degrades at low momentum, further supporting the use of momentum thresholds in the selection.

\begin{figure*}[h!]
    \centering
    \begin{subfigure}[t]{0.5\textwidth}
        \centering
        \includegraphics[width=0.95\textwidth]{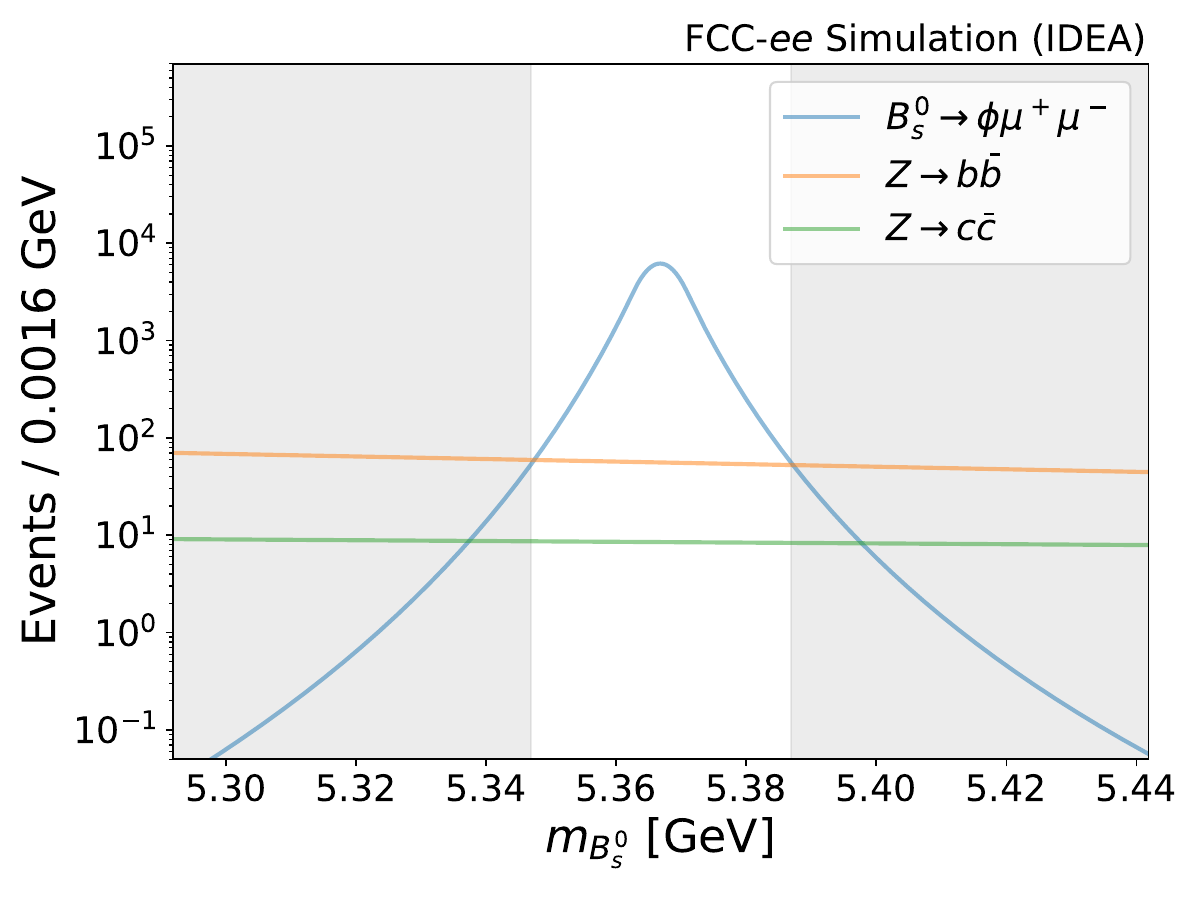}
    \end{subfigure}%
    \caption{The fitted distributions of the reconstructed ${B_s^0}$ mass, according to Appendix~\ref{app:mBs}.}
    \label{fig:mBs}
\end{figure*}

The reconstructed invariant mass information is shown in Figures~\ref{fig:reson} and~\ref{fig:mBs}. A clear $\phi$ peak is visible in all samples. The tails are dominated by random combinatorics from Kaons. In the $q^2$ distribution, which is equivalent to the squared dimuon invariant mass distribution, the backgrounds exhibit resonances around the $m_{J/\psi}^2$ and $m_{\psi}^2$ mass regions.\footnote{The resonance around $m_\phi^2$ is not visible because the Br$(\phi\to \mu^+\mu^-)=2.85\times 10^{-4}$ is small compared to Br$(J/\psi\to \mu^+\mu^-)=5.961\%$ and Br$(\psi\to \mu^+\mu^-)=8.0 \times 10^{-3}$.} Note that the observed peaks are not solely due to resonant backgrounds, but also include contributions from other decays that produce intermediate states such as $\phi$, $J/\psi$, or $\psi$. For example, a $\phi$ meson can originate from a $D_s$ decay.

Due to the limited number of simulated events and the efficiency of the selection criteria, only a few events remain after their application. To avoid the effects of statistical fluctuations in the simulation, we fit the $m_{B_s^0}$ distribution using the samples after the $q^2$ selection requirement. The fitted results are shown in Figure~\ref{fig:mBs}, with the signal and backgrounds modeled using double-sided Crystal Ball and exponential distributions, respectively. Details of the fitting procedure are provided in Appendix~\ref{app:mBs}. The total estimated signal yield is $3.93 \times 10^4$. For the backgrounds, the estimated yields are $1.39 \times 10^3$ from $Z \rightarrow b\bar{b}$ and $2.13 \times 10^2$ from $Z \rightarrow c\bar{c}$ (see Table~\ref{tab:cutFlow}).

\section{Results\label{sec:results}}

The time-dependent differential decay rate can be written as: 
\begin{equation}
    \frac{d\Gamma(t)}{dq^2} = \frac{e^{-\Gamma t}}{2}\Bigg[C_1^\phi\cos\big(\Delta m\,t\big) + C_2^\phi\sin\big(\Delta m\,t\big) + C_3^\phi\cosh\Big(\frac{\Delta \Gamma t}{2}\Big) + C_4^\phi \sinh\Big(\frac{\Delta \Gamma t}{2}\Big)\Bigg]\,, \label{eq:tdiff}
\end{equation}
where $C^\phi_i$ are functions of the helicity amplitudes. More details can be found in Section~\ref{sec:inter}. 

Due to the finite decay time resolution, the measured decay time is smeared. As a consequence, the observed oscillations are diluted by a given factor~\cite{Moser:1996xf}:
\begin{align}
    \mathcal{D}_{\rm time} = e^{-\frac{1}{2}\Delta m_s^2 \sigma_t^2} \ ,
\end{align}
where $\sigma_t$ is the time resolution. This value is dominated by the secondary vertex resolution. Considering the excellent vertexing capabilities of future detector concepts, we estimate the secondary vertex resolution to be $\mathcal{O}(10~\mu{\rm m})$, as shown in Figure~\ref{fig:vertex}. Given the average decay length of the $B_s^0$ meson is approximately 2.8\,mm at the $Z$-pole, the estimated decay length resolution is about $3.57 \times 10^{-3}$\,mm.\footnote{{The time resolution depends on both the decay length resolution and the momentum resolution, though the momentum resolution is neglected here. For further discussion, please refer to~\cite{Lucchini:2020bac}. Additionally, the decay length resolution is influenced by both the primary and secondary vertex resolutions, with the secondary vertex resolution being the dominant factor. As a result, we neglect the contribution of the primary vertex resolution here.}} Based on the mean lifetime of the $B_s^0$ meson, approximately 1.529\,ps, we estimate the time resolution to be $\sigma_t \approx 5.5 \times 10^{-3}$\,ps. This results in a time dilution factor of $\mathcal{D}_{\rm time}\approx 0.995$. Thanks to the excellent vertexing capabilities expected at future $Z$-factories, even if the vertex resolution is degraded by a factor of 2, the time dilution factor would experience only a 1.5\% relative reduction. This supports the robustness of the following study on time-dependent measurements. It is important to note that, in practice, time-dependent measurements are also influenced by the detector acceptance, which varies with the decay length and the boost. In this study, we focus exclusively on the decay time region between 1 and 8\,ps to suppress prompt background contributions and avoid regions with poor statistical precision or reconstruction inefficiencies.

Measurements of $C\!P$-asymmetry require knowledge of the initial flavor of the neutral meson. Modern flavor tagging algorithms exploit various features to identify the initial flavor, including the decay of the opposite-side $b$-quark, the fragmentation of the signal-side $b$-quark, and the overall event topology. Imperfect flavor tagging effectively reduces the sample size by the factor known as tagging power (see Appendix~\ref{app:acpUncert}). Table~\ref{tab:tagPower} shows the tagging powers from different experiments.\footnote{This represents the typical tagging power; however, the actual value depends on the decay mode.} Improvements in flavor tagging are crucial for the success of the $C\!P$-asymmetry measurements in future machines.

\begin{table}[h!]
    \centering
    \begin{tabular}{cccccc}
        \hline  
        \hline
      & LEP & Belle~II & BaBar & LHCb & CMS \\
        \hline
    $P_{\rm tag}$ & $25-30\%$ & $30\%$ & $30\%$ & $6\%$ & $5-10\%$ \\
        \hline  
        \hline
    \end{tabular}
    \caption{Tagging power from current experiments~\cite{Aleksan:2021gii, Belle-II:2021zvj, CMS:2020efq}.}
    \label{tab:tagPower}
\end{table}

In the following subsections, we will discuss the potential measurements with $B_s^0\to\phi\mu^+\mu^-$ decays that can be performed at future $Z$-factories. These measurements can be divided into two major categories: untagged measurements, where tagging the $B^0_s$ is not necessary, and tagged measurements, where tagging is required. The relevant parameters for these measurements are summarized in Table~\ref{tab:Bsparam}. 

As the $\phi\mu^+\mu^-$ final state is a mixture of $C\!P$-eigenstates, the final amplitudes are polarized and have different decay time dependences. 
These polarization amplitudes have to be statistically separated to measure the $C\!P$-violating phase. 
This can be achieved by measuring the angular-dependent decay rate. 
However, since the primary advantage of FCC-$ee$ lies in its time-dependent measurement capabilities rather than angular analysis, we integrate over the polarized final states in this study.
As a result, we forgo the ability to disentangle the individual contributions of the polarized amplitudes in order to extract the $C\!P$-violating phase from the time-dependent asymmetry. However, because the decay is dominated by the $C\!P$-even amplitude throughout the entire $q^2$ range~\cite{LHCb:2021xxq}, there is no full cancellation or phase flip between $C\!P$-even and $C\!P$-odd contributions. This ensures that integrating over polarization does not eliminate potential $C\!P$-violating effects in the time-dependent analysis.

\begin{table}[h!]
    \centering
    \begin{tabular}{cccccc}
        \hline  
        \hline
        Parameter & $\Delta m_s$ & $\Delta\Gamma_s $ & $1/\Gamma_s$ & $\mathcal{D}_{\rm time}$ & $P_{\rm tag}$  \\
        \hline
        Value & 17.765~ps$^{-1}$ & 0.084~ps$^{-1}$ & 1.520~ps & 0.995 & 0.3  \\
        \hline  
        \hline
    \end{tabular}
    \caption{Parameter choices assumed to perform the measurements in this study. The $B_s^0$ related parameters are taken from~\cite{Workman:2022ynf}.}
    \label{tab:Bsparam}
\end{table}

\subsection{Untagged Measurements}
Here we discuss two untagged measurements. We start with projections on the branching ratio, which does not require knowledge of the decay time. Then we show the sensitivity of the untagged time-dependent decay rate measurement. 

\paragraph{Branching Ratio} 
We estimate the statistical uncertainty on Br($B_s^0\to\phi\mu^+\mu^-$) as $\sqrt{s+b}/s$, where the signal ($s$) and background ($b$) yields can be extracted from Table~\ref{tab:cutFlow}. We estimate that the branching ratio can be known with a relative precision of $0.515\%$, which is roughly a factor of 5 better than the LHCb statistical uncertainty~\cite{LHCb:2021zwz}. The low-$q^2$ ($q^2\in[1.1, 6.0]$~GeV$^{2}$) and high-$q^2$ ($q^2\geq 15$~GeV$^{2}$) ranges are sensitive to different EFT operators (see Section~\ref{sec:inter}). The relative precisions of these two regions are $0.810\%$ and $1.577\%$, respectively.

\paragraph{Untagged Time-dependent Decay Rate} 

\begin{figure*}[h!]
    \centering
    \begin{subfigure}[t]{0.5\textwidth}
        \centering
        \includegraphics[width=0.95\textwidth]{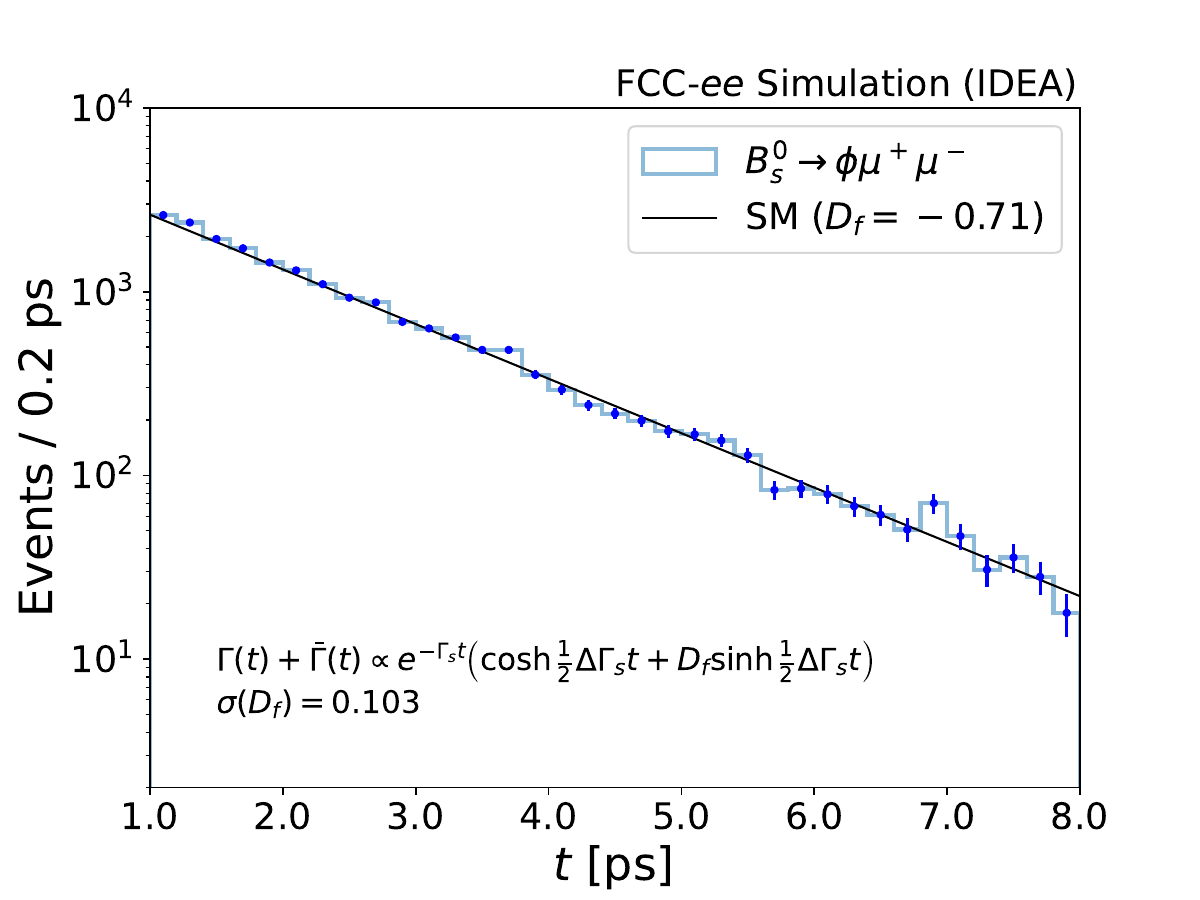}
    \end{subfigure}%
    \caption{Distribution and fit of the untagged time-dependent decay rate.}
    \label{fig:Df}
\end{figure*}

According to Eq.~\eqref{eq:tdiff}, one can write the untagged time-dependent decay rate as:
\begin{align}
      \Gamma_{B_s^0\to \phi\mu^+\mu^-}(t) + \Gamma_{\bar{B}_s^0\to \phi\mu^+\mu^-}(t) \propto e^{-\Gamma_s t}\left(\cosh{\frac{1}{2}\Delta\Gamma_s t}+D_f \sinh{\frac{1}{2}\Delta\Gamma_s t}\right) \ .
      \label{eq:sumRate}
\end{align}
Notably, unlike in $B^0$ studies where $\Delta\Gamma_d\approx 0$, making $D_f$ immeasurable due to ${\sinh{\frac{1}{2}\Delta\Gamma_d t}\approx 0}$, in the case of the $B_s^0$ meson, which is the focus of our study, the fact that $\Delta\Gamma_s\approx 0.085$~ps$^{-1}$ makes the measurement of $D_f$ feasible. In Figure~\ref{fig:Df}, we show the simulated untagged decay time distribution. Our fit of Eq.~\eqref{eq:sumRate} gives us a statistical uncertainty on $D_{f}$ of $0.103$  with a value of $D_f=-0.71$,\footnote{Given the absence of experimental measurements for $D_f$ and the theoretical uncertainty in its calculation, we vary the value of $D_f$ between $-0.6$ and $-0.8$, with the uncertainty remaining within the range $0.103$ to $0.104$. Notably, the uncertainty of $D_f$ is not sensitive to the value of $D_f$.} as calculated in Section~\ref{sec:inter}.

\subsection{Tagged Measurements}
Time-integrated and time-dependent measurements are also possible in the tagged cases. Specifically, we can measure both the time-integrated and time-dependent $C\!P$ asymmetry: $\langle A_{C\!P}\rangle $ and $A_{C\!P}(t)$, where the $C\!P$ asymmetry is defined by: 
\begin{align}
    A_{C\!P}(t)&\equiv \frac{\Gamma_{B_s^0\to \phi\mu^+\mu^-}(t) - \Gamma_{\bar{B}_s^0\to \phi\mu^+\mu^-}(t)}{\Gamma_{B_s^0\to \phi\mu^+\mu^-}(t) + \Gamma_{\bar{B}_s^0\to \phi\mu^+\mu^-}(t)} \,.
    \label{eq:acp}
\end{align}

\begin{figure*}[h!]
    \centering
    \begin{subfigure}[t]{0.5\textwidth}
        \centering
        \includegraphics[width=1\textwidth]{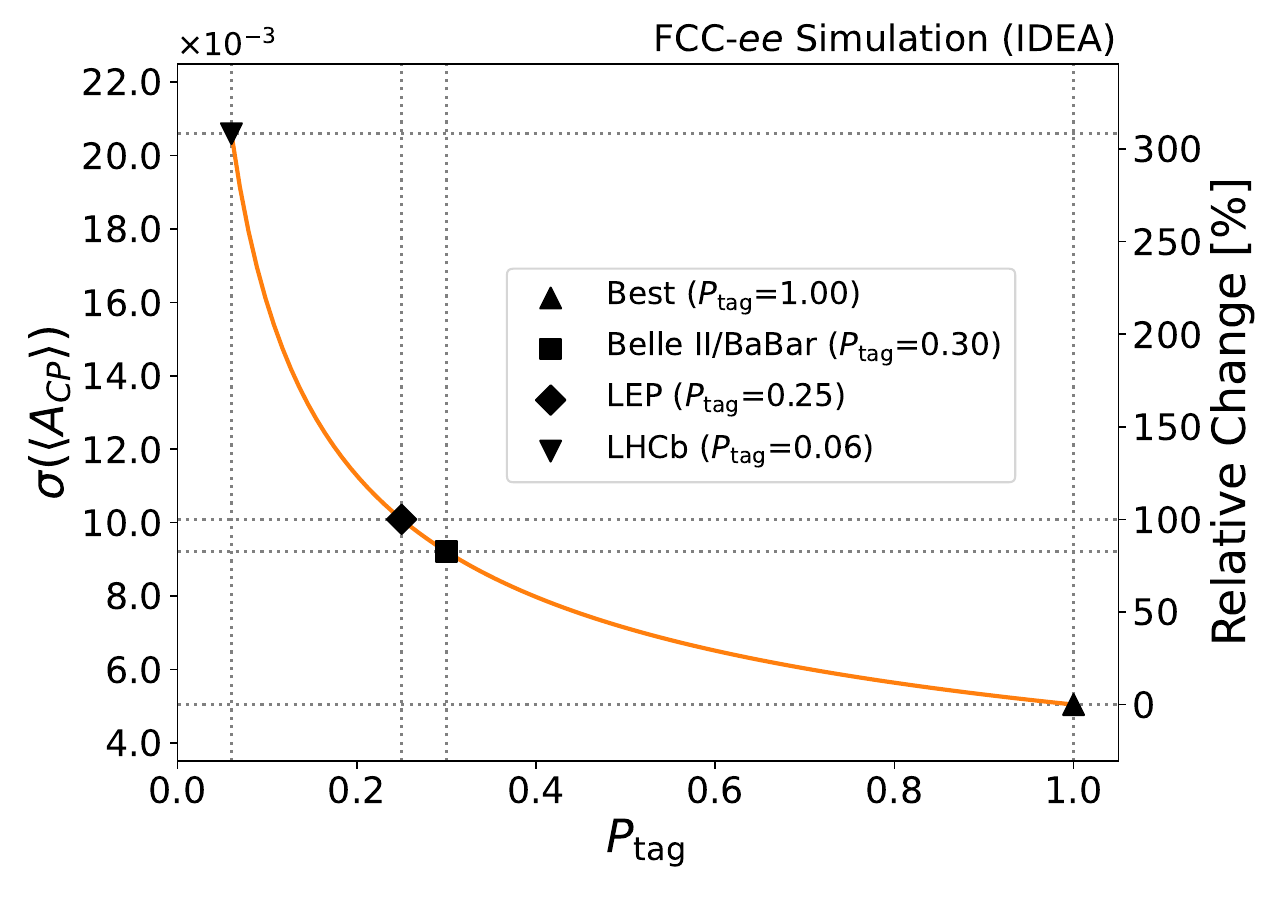}
    \end{subfigure}%
    \caption{Absolute statistical uncertainty of time-integrated $C\!P$-asymmetry measurement as a function of tagging power. The relative change in statistical uncertainty with respect to the best case scenario ($P_{\rm tag}=1$) is shown on the right-hand side.}
    \label{fig:acp_best}
\end{figure*}

\paragraph{Time-integrated $C\!P$-asymmetry}
The integrated $C\!P$-asymmetry is defined as
\begin{align}
    \langle A_{C\!P}\rangle = \frac{\Gamma_{B_s^0\to \phi\mu^+\mu^-} - \Gamma_{\bar{B}_s^0\to \phi\mu^+\mu^-}}{\Gamma_{B_s^0\to \phi\mu^+\mu^-} + \Gamma_{\bar{B}_s^0\to \phi\mu^+\mu^-}}\ .
    \label{eq:acpTimeGeneral}
\end{align}
The statistical uncertainty on $\langle A_{C\!P}\rangle$ can be estimated by $1/\sqrt{P_{\rm tag}\cdot s}$, where $P_{\rm tag}$ is the tagging power, and $s$ is the total signal yield of $B_s^0\to\phi\mu^+\mu^-$, assuming the background is statistically subtracted. A conservative estimate gives an uncertainty of $\sigma\left(\langle A_{C\!P}\rangle\right)=9.21\times 10^{-3}$ with tagging power $P_{\rm tag} = 0.3$, matching that of Belle~II or BaBar. Figure~\ref{fig:acp_best} shows the expected $\sigma\left(\langle A_{C\!P}\rangle\right)$ as a function of $P_{\rm tag}$, where different benchmarks are also marked as a reference.

\begin{figure*}[h!]
    \centering
    \begin{subfigure}[t]{0.5\textwidth}
        \centering
        \includegraphics[width=0.95\textwidth]{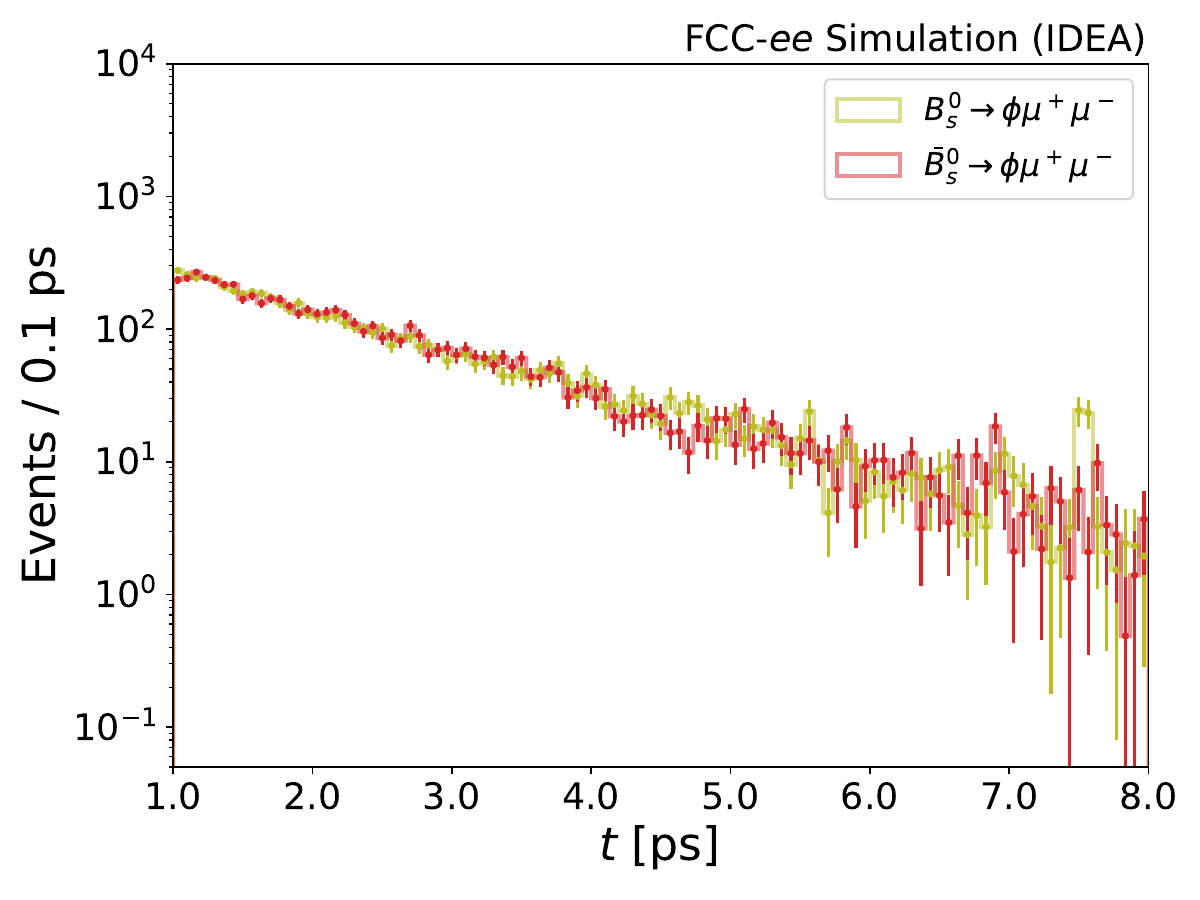}
    \end{subfigure}%
    ~
    \begin{subfigure}[t]{0.5\textwidth}
        \centering
        \includegraphics[width=0.95\textwidth]{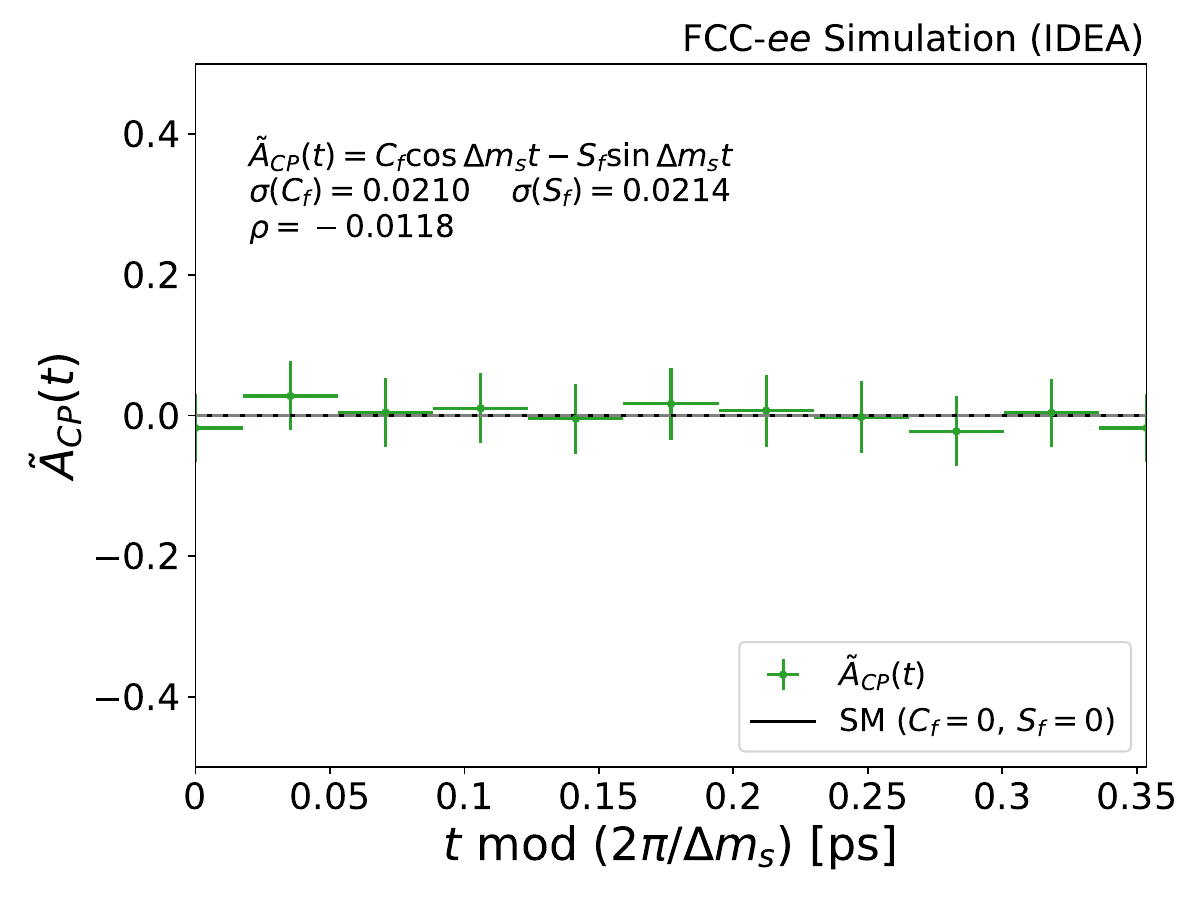}
    \end{subfigure}%
    \caption{\textbf{Left}: Distributions of the tagged time-dependent decay rate measurement of $B_s^0$ and $\bar{B}_s^0$ decays, in the SM. \textbf{Right}: The modified $C\!P$-asymmetry measurement with fit, in the SM.}
    \label{fig:acp}
\end{figure*}

\paragraph{Time-dependent $C\!P$-asymmetry}
With tagging present, one can write the time-dependent $C\!P$-asymmetry, using Eq.~\eqref{eq:tdiff}, as:
\begin{align}
    A_{C\!P}(t) = \frac{C_f \cos{\Delta m_s t} - S_f \sin{\Delta m_s t}}{\cosh{\frac{1}{2}\Delta\Gamma_s t}+D_f\sinh{\frac{1}{2}\Delta\Gamma_s t}}\ .
    \label{eq:acpTime}
\end{align}
In addition to $D_f$, $C_f$ and $S_f$ become accessible in the fully tagged time-dependent measurement. This allows for directly fitting the $C_f$ and $S_f$ parameters from $A_{C\!P}(t)$. Here we fix $D_f$ to the SM value in the fit. Note that, in this study, we do not attempt to unfold the values of the $C\!P$-violating phase, strong phases, and amplitudes. First, the SM predicts $C_f = S_f = 0$, leading to a degeneracy in the value of $D_f$ in this context. Second, the $D_f$ parameter is better constrained in the untagged time-dependent decay rate measurement, where we expect the precision to reach $\mathcal{O}(0.1)$, as discussed above. Even if a small NP contribution with $\delta D_f^{\rm NP}\sim \mathcal{O}(0.1)$ exists, such that it remains undetected in the untagged time-dependent decay rate measurement, this small effect is negligible here, as $\sinh{\frac{1}{2}\Delta\Gamma_s t}\lesssim\mathcal{O}(0.1)$. Figure~\ref{fig:acp} shows the decay rates of the $B_s^0\to\phi\mu^+\mu^-$ and $\bar{B}_s^0\to\phi\mu^+\mu^-$ distributions. Note that the bin size of this measurement is smaller than that of the untagged measurement (see Figure~\ref{fig:Df}), and hence introduces a larger statistical fluctuation.\footnote{In a real analysis, an unbinned fit can be performed.} This finer binning is necessary to resolve the oscillation in the data, as the sampling frequency must exceed twice the oscillation frequency. The fit results show that we can achieve uncertainties of $\sigma(C_f)=0.0210$ and $\sigma(S_f)=0.0214$, with a small correlation of $\rho=-0.0118$. To reduce the statistical uncertainty of the bins, we can also reproduce a one-period oscillation with a modified asymmetry function defined as:
\begin{align}
    \widetilde{A}_{C\!P}(t) &= A_{C\!P}(t)\times \left(\cosh{\frac{1}{2}\Delta\Gamma_s t}+D_f\sinh{\frac{1}{2}\Delta\Gamma_s t}\right)\nonumber\\
    &= C_f \cos{\Delta m_s t} - S_f \sin{\Delta m_s t}\ ,
\end{align}
which is also shown in Figure~\ref{fig:acp}, where the statistical fluctuation is significantly smaller. 

\begin{figure*}[h!]
    \centering
    \begin{subfigure}[t]{0.5\textwidth}
        \centering
        \includegraphics[width=0.95\textwidth]{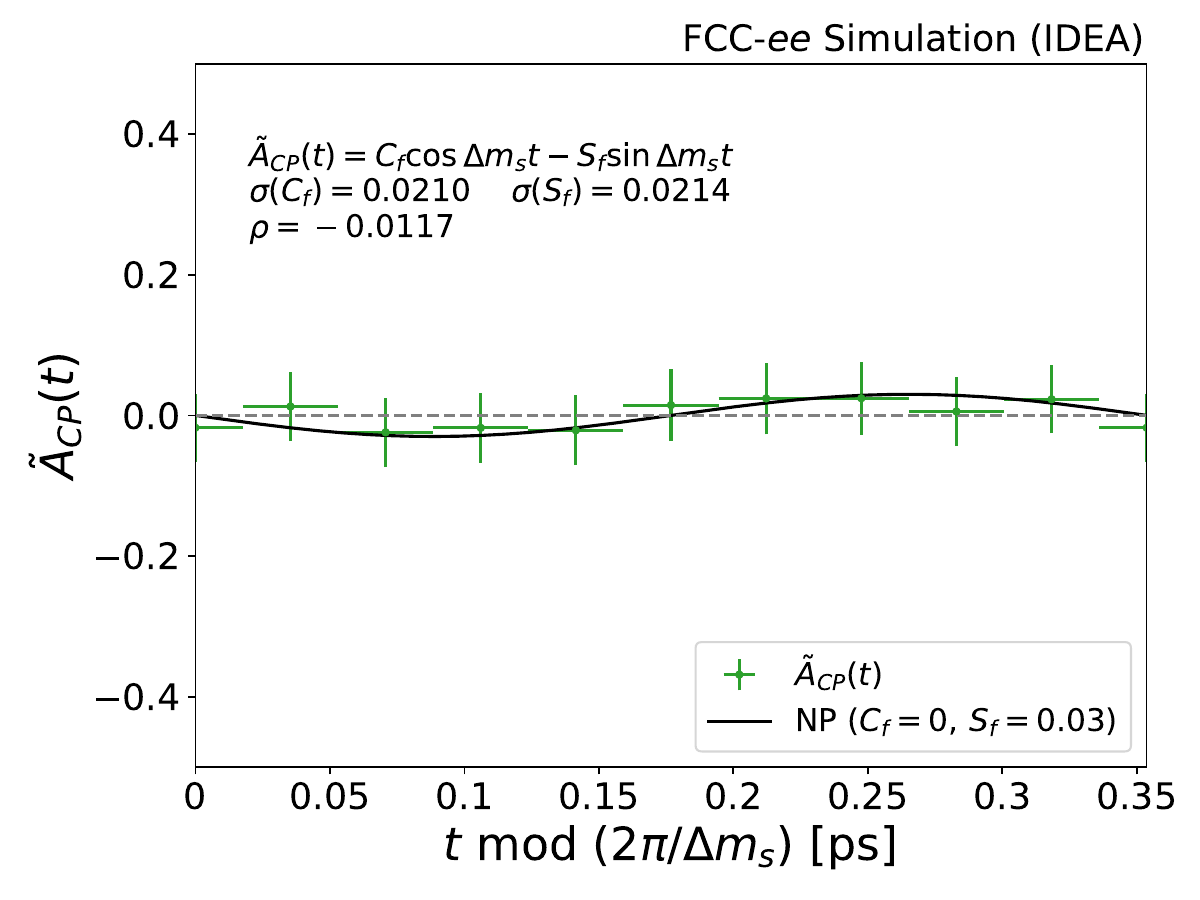}
    \end{subfigure}%
    ~ 
    \begin{subfigure}[t]{0.5\textwidth}
        \centering
        \includegraphics[width=0.95\textwidth]{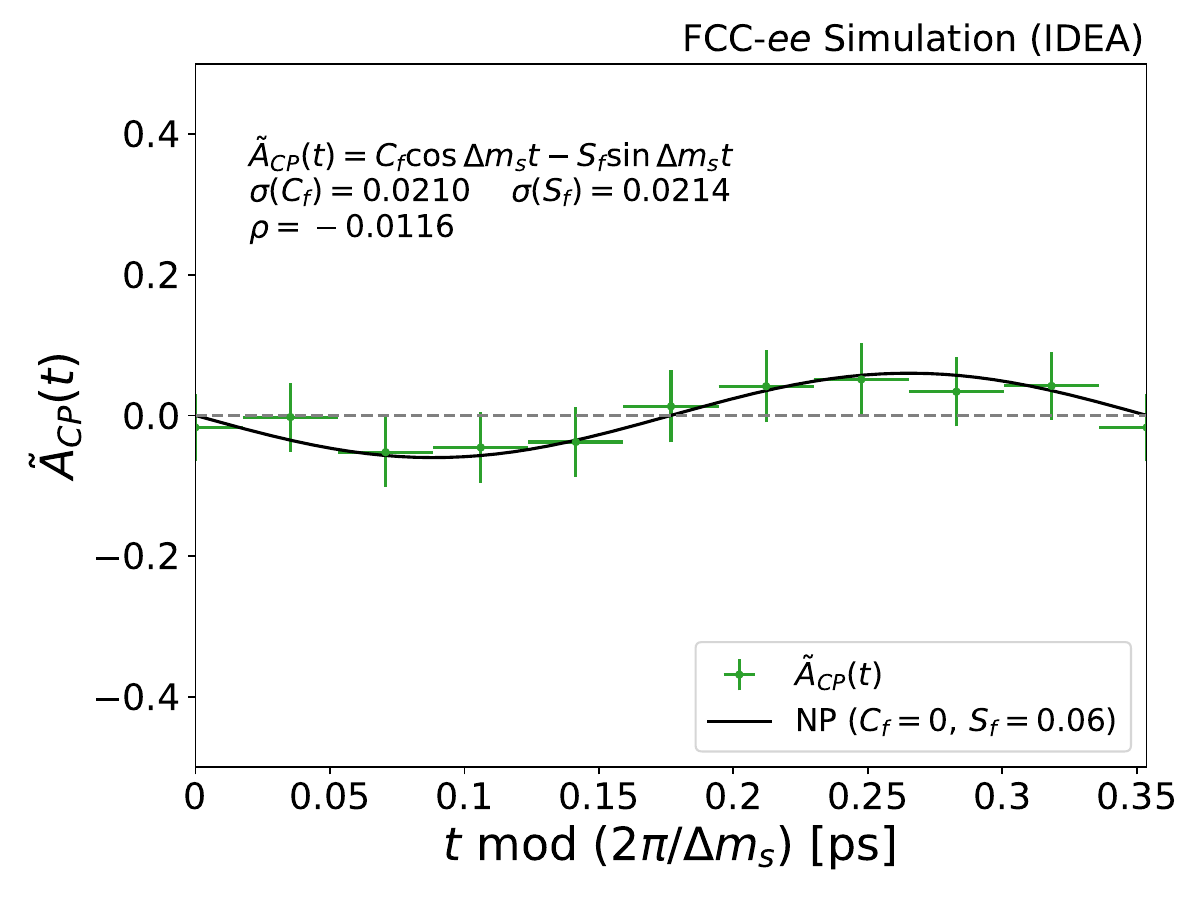}
    \end{subfigure}
    
    \begin{subfigure}[t]{0.5\textwidth}
        \centering
        \includegraphics[width=0.95\textwidth]{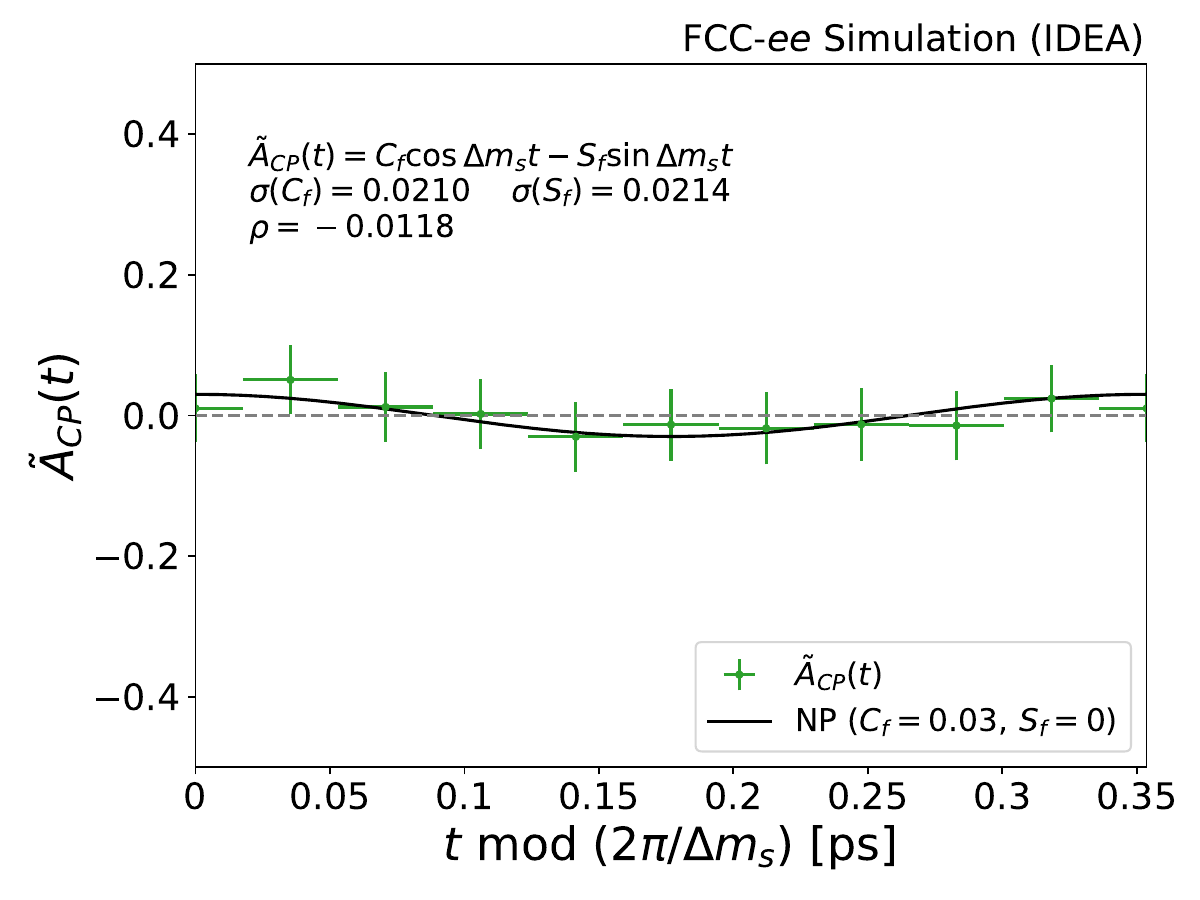}
    \end{subfigure}%
    ~ 
    \begin{subfigure}[t]{0.5\textwidth}
        \centering
        \includegraphics[width=0.95\textwidth]{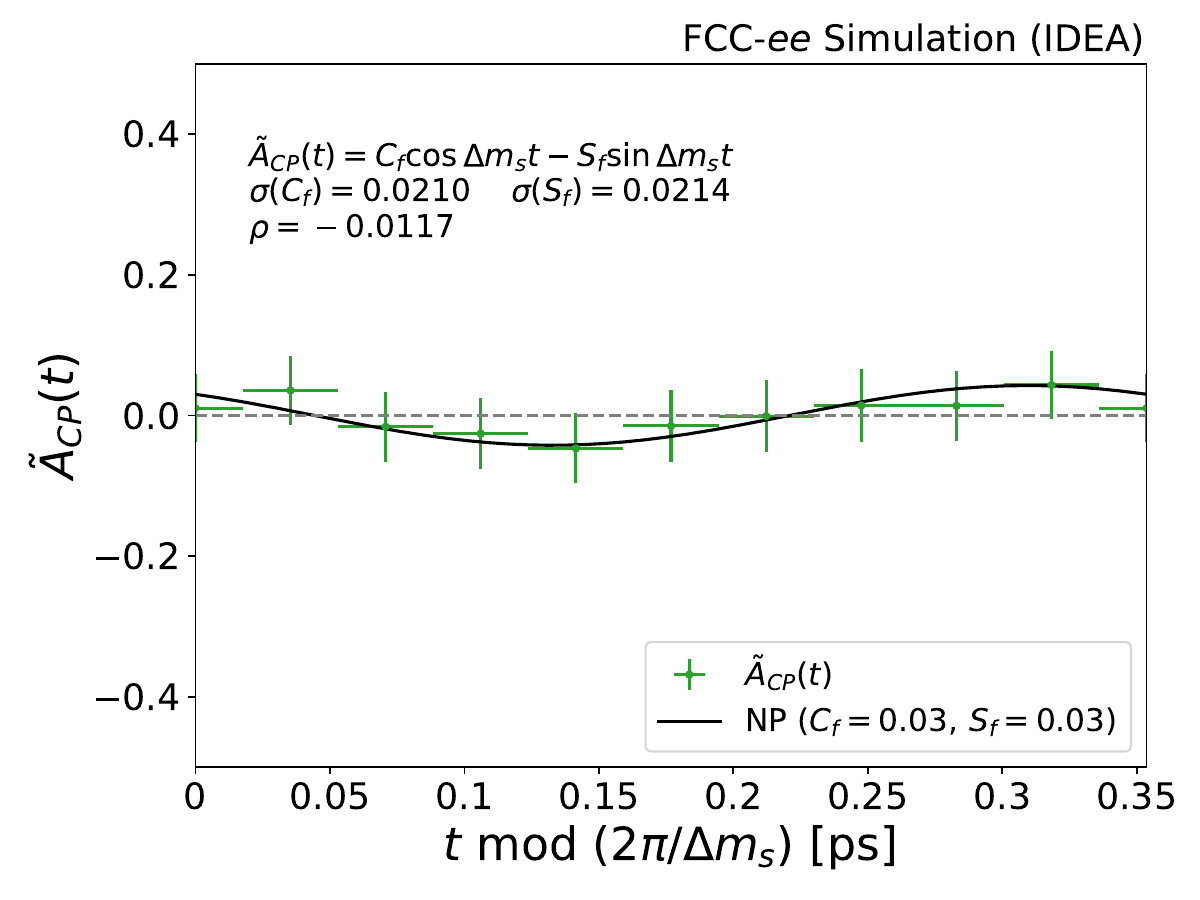}
    \end{subfigure}%

    \caption{Four examples of NP scenarios are shown: modified with ${\delta S_f^{\rm NP}=0.03}$, ${\delta S_f^{\rm NP}=0.06}$, ${\delta C_f^{\rm NP}=0.03}$, and with both ${\delta S_f^{\rm NP}=0.03}$ and ${\delta C_f^{\rm NP}=0.03}$, respectively.}
    \label{fig:acpNP}
\end{figure*}

Furthermore, we investigate the effects of NP on the time-dependent $C\!P$ asymmetry $A_{CP}(t)$. We consider scenarios in which NP introduces a new phase, modifying the observables as 
\begin{align}
C_f = C_f^{\mathrm{SM}} + \delta C_f^{\mathrm{NP}}, \quad S_f = S_f^{\mathrm{SM}} + \delta S_f^{\mathrm{NP}}\ .
\end{align}
In the SM, both $C_f^{\mathrm{SM}}$ and $S_f^{\mathrm{SM}}$ are expected to be small or vanish; thus, any significant deviation from zero in these observables would point to potential NP contributions. We select $(\delta C_f^{\rm NP}, \delta S_f^{\rm NP})=(0, 0.03),\ (0, 0.06),\ (0.03, 0)$ and $(0.03, 0.03)$, as test ranges for possible NP contributions (see Figure~\ref{fig:acpNP}).

\begin{figure*}[h!]
    \centering
    \begin{subfigure}[t]{0.6\textwidth}
        \centering
        \includegraphics[width=0.95\textwidth]{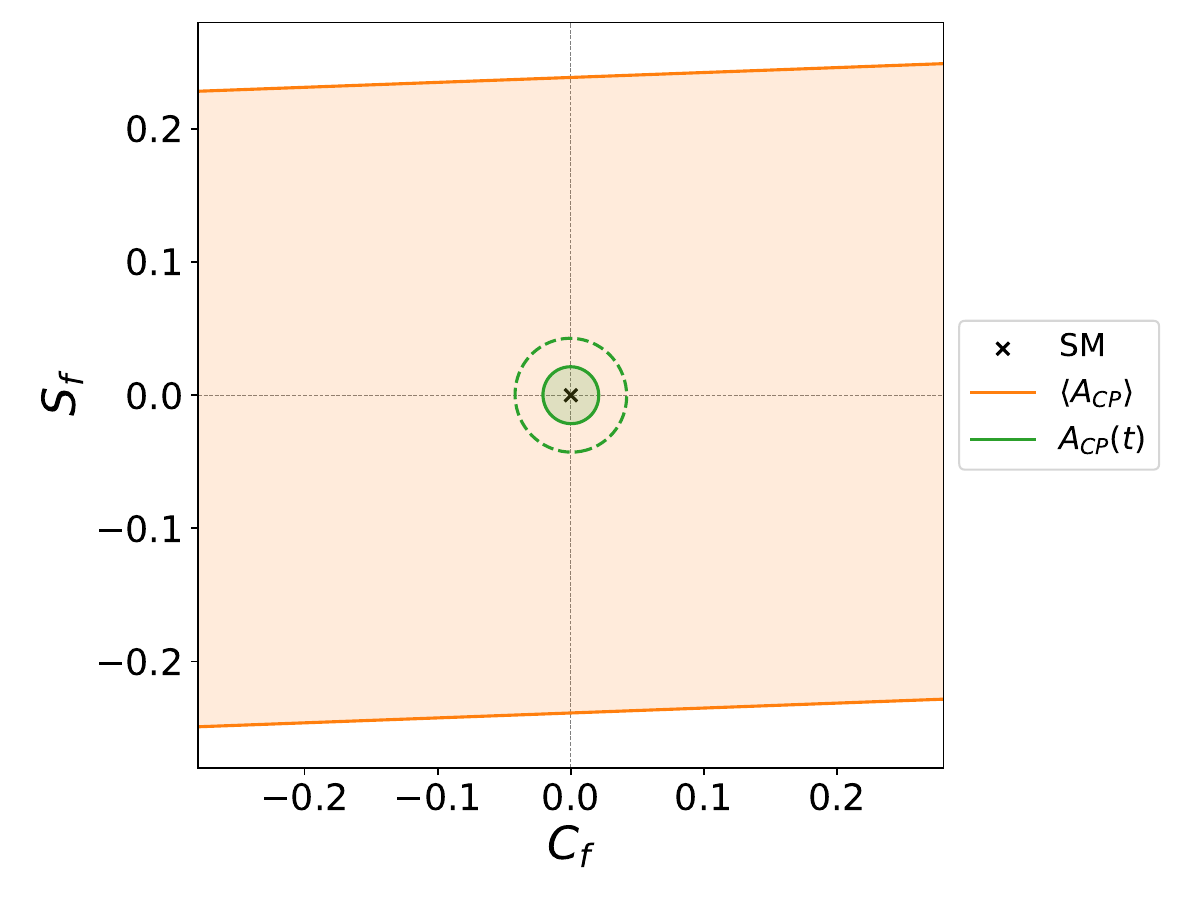}
    \end{subfigure}%
    \caption{Constraints on the $S_f$-$C_f$ plane from various measurements, including the time-dependent decay rate, time-integrated $C\!P$ asymmetry, and time-dependent $C\!P$ asymmetry. The solid line represents the $1\sigma$ constraint, while the dashed line indicates the $2\sigma$ constraint.}
    \label{fig:summary}
\end{figure*}

Notably, based on Eq.~\eqref{eq:acpTimeGeneral}, the integration yields:
\begin{align}
    \langle A_{C\!P}\rangle = \frac{4\Gamma_s^2 - \Delta \Gamma_s^2}{\Gamma_s^2 + \Delta m_s^2 }\,\frac{C_f \Gamma_s - S_f \Delta m_s }{4\Gamma_s + 2 D_f \Delta \Gamma_s} \ , 
\end{align}
which shows that the measurement of $\langle A_{C\!P}\rangle$ can be represented as a function of $C_f$ and $S_f$. Due to the fact that $\Delta m_s > \Gamma_s$, the sensitivity to $S_f$ will be better than that of $C_f$. A summary of the sensitivities of the measurements in the $C_f$-$S_f$ plane is shown in Figure~\ref{fig:summary}, when fixing $D_f$ to the SM value. One finds that the time-dependent $C\!P$-asymmetry measurement is better in measuring $C_f$ and $S_f$ from the tagged time-dependent measurements, which are approximately an order of magnitude more precise than the time-integrated measurement. This highlights the significance of performing such measurements at the FCC-$ee$, which provides opportunities to directly measure $C_f$ and $S_f$ instead of inferring them from other measurements.

A summary of the measurements discussed in this section is presented in Table~\ref{tab:summary}. The precision for the branching ratio can reach $<\mathcal{O}(1\%)$, and for $\langle A_{C\!P}\rangle$ it can reach $\mathcal{O}(10^{-2})$. The precision for $D_f$ can reach $\mathcal{O}(10^{-1})$, while for $C_f$ and $S_f$ it can reach $\mathcal{O}(10^{-2})$.

\begin{table}[h!]
    \centering
    \begin{tabular}{ccccccc}
        \hline \hline\\[-12pt]
        $\frac{\sigma ( {\rm Br})}{\rm Br}$ & $\frac{\sigma ({\rm Br}^{q^2\in [1.1, 6]})}{{\rm Br}^{q^2\in [1.1, 6]}}$ & $\frac{\sigma ( {\rm Br}^{q^2 \geq 15})}{{\rm Br}^{q^2 \geq 15}}$ & $\sigma (D_{f})$ & $\sigma ( \langle A_{ C\!P}\rangle )$ & $\sigma ( C_f)$ & $\sigma ( S_f)$\\
        [7pt]\hline
        $0.515\%$ & $0.810\%$ & $1.577\%$ & $0.103$ & $9.21\times 10^{-3}$ & $0.0210$ & $0.0214$\\
        \hline \hline
    \end{tabular}
    \caption{Summary of the experimental precision of the observables. }
    \label{tab:summary}
\end{table}

\section{Interpretation~\label{sec:inter}}

\subsection{Effective Field Theory}
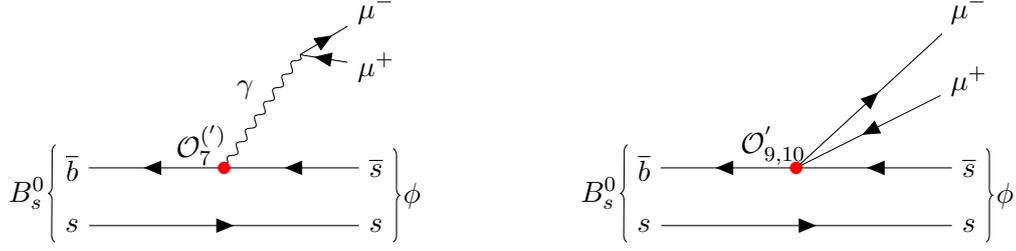
\begin{figure}[t!]
\centering  
    \begin{subfigure}[b!]{0.49\textwidth}
        \centering
        \begin{tikzpicture}
        \begin{feynman}
        \vertex (a1) {\(\overline b\)};
        \vertex[right=2cm of a1, red, dot] (a2) {};
        \vertex (e1) at ($(a2)+(-0.3, 0.3)$) {\(\mathcal{O}^{(')}_7\)};
        \vertex[right=2cm of a2] (a4) {\(\overline s\)};
        \vertex (c2) at ($(a4)+(-1,1.5)$);   
        \vertex[below=2em of a1] (b1) {\(s\)};
        \vertex[below=2em of a4] (b2) {\(s\)};
        \vertex at ($(a1) + (1.4+0.6, 0.6)$) (d);
        \vertex[above=3.5em of a4] (c1) {\(\mu^+\)};
        \vertex[above=2em of c1] (c3) {\(\mu^-\)};
        
        \diagram* {
        (a4) -- [fermion] (a2) -- [fermion] (a1),
        (b1) -- [fermion] (b2),
        (c1) -- [fermion] (c2) -- [fermion] (c3),
        (a2) -- [boson, edge label=\(\gamma \)] (c2),
        };
        \draw [decoration={brace}, decorate] (b1.south west) -- (a1.north west)
        node [pos=0.5, left] {\(B_s^0\)};
        \draw [decoration={brace}, decorate] (a4.north east) -- (b2.south east)
        node [pos=0.5, right] {\(\phi\)};
        \end{feynman}
        \end{tikzpicture}  
    \end{subfigure}%
    ~
    \begin{subfigure}[b!]{0.49\textwidth}
        \centering
        \begin{tikzpicture}
        \begin{feynman}
        \vertex (a1) {\(\overline b\)};
        \vertex[right=2cm of a1, red, dot] (a2) {};
        \vertex (e1) at ($(a2)+(-0.3, 0.3)$) {\(\mathcal{O}^{'}_{9,10}\)};
        \vertex[right=1.45cm of a3] (a4) {\(\overline s\)};
        \vertex[below=2em of a1] (b1) {\(s\)};
        \vertex[below=2em of a4] (b2) {\(s\)};
        
        \vertex[above=3.em of a4] (c1) {\(\mu^+\)};
        \vertex[above=2.5em of c1] (c3) {\(\mu^-\)};
       
        \diagram* {
        (a4) -- [fermion] (a2) -- [fermion] (a1),
        (b1) -- [fermion] (b2),
        (a2) -- [fermion] (c3),
        (c1) -- [fermion] (a2),
        };
        \draw [decoration={brace}, decorate] (b1.south west) -- (a1.north west)
        node [pos=0.5, left] {\(B_s^0\)};
        \draw [decoration={brace}, decorate] (a4.north east) -- (b2.south east)
        node [pos=0.5, right] {\(\phi\)};
        \end{feynman}
        \end{tikzpicture}
    \end{subfigure}
    \caption{The leading order WET Feynman diagrams of $B_s^0\to\phi\mu^+\mu^-$. \textbf{Left}: Contribution from the radiative $\mathcal{O}_7$ operator, where $\gamma$ decays to $\mu^+\mu^-$. \textbf{Right}: Contribution from $\mathcal{O}_{9}$ and $\mathcal{O}_{10}$ operators. }
    \label{fig:EFTFD}
\end{figure}

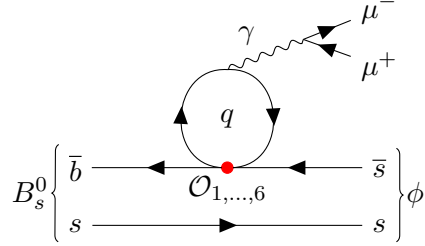
\begin{figure}[t]
    \centering
        \begin{tikzpicture}
        \begin{feynman}
        \vertex (a1) {\(\overline b\)};
        \vertex[right=2cm of a1, red, dot] (a2) {};
        \vertex (e1) at ($(a2)+(0.0, -0.3)$) {\(\mathcal{O}_{1,\dots,6}\)};
        \vertex[right=2cm of a2] (a4) {\(\overline s\)};
        \vertex (c2) at ($(a4)+(-1,1.7)$);   
        \vertex (d1) at ($(a2)+(0, 1.3)$);
        \vertex (e2) at ($(a2) + (0, 0.65)$) {\(q\)};
        \vertex[below=2em of a1] (b1) {\(s\)};
        \vertex[below=2em of a4] (b2) {\(s\)};
        \vertex at ($(a1) + (1.4+0.6, 0.6)$) (d);
        \vertex[above=3.5em of a4] (c1) {\(\mu^+\)};
        \vertex[above=2em of c1] (c3) {\(\mu^-\)};
        
        \diagram* {
        (a4) -- [fermion] (a2) -- [fermion] (a1),
        (b1) -- [fermion] (b2),
        (c1) -- [fermion] (c2) -- [fermion] (c3),
        (a2) -- [fermion, half left] (d1) -- [fermion, half left] (a2),
        (d1) -- [boson, edge label=\(\gamma \)] (c2),
        };
        \draw [decoration={brace}, decorate] (b1.south west) -- (a1.north west)
        node [pos=0.5, left] {\(B_s^0\)};
        \draw [decoration={brace}, decorate] (a4.north east) -- (b2.south east)
        node [pos=0.5, right] {\(\phi\)};
        \end{feynman}
        \end{tikzpicture}  
        \caption{Contribution to $C_{7,9}^{\rm eff}$ from one-loop matrix elements of current-current operators, $\mathcal{O}_{1,\dots,6}$. In the diagram, $q=u,d,s,c,b$.}
        \label{fig:currcurr}
\end{figure}

The $b\to s\mu^+\mu^-$ transition is described in the WET, after integrating out heavy particles at the electroweak scale, by the Hamiltonian
\begin{align}
    \mathcal{H}^{\rm eff}&\supset-\frac{4G_F V_{tb}V^\ast_{ts}}{\sqrt{2}}\left(\sum_{i=1}^{8}\big(C_i\mathcal{O}_i + C'_i {\mathcal{O}_i}'\big) + \sum_{i=9}^{10}\big(C_i\mathcal{O}^\mu_i + C'_i {\mathcal{O}^\mu_i}'\big)  \right) + {\rm h.c.}\ .
    \label{eq:EFT}
\end{align}
Of the twenty operators in Eq.~\eqref{eq:EFT}, six give direct contributions to the $B_s^0\to\phi\mu^+\mu^-$ decay, illustrated in Figure~\ref{fig:EFTFD}, namely
\begin{equation}\begin{split}
    &\mathcal{O}_7 = \frac{e}{(4\pi)^2} m_b[\bar{s}\sigma^{\mu\nu} P_R b]F_{\mu\nu}\,,\quad \mathcal{O}_{9(10)}^\mu = \frac{e^2}{(4\pi)^2}[\bar{s}\gamma^\mu P_L b][\bar{\mu}\gamma_\mu(\gamma_5)\mu] \ ,\\[0.5em]
    &\mathcal{O}'_7 = \frac{e}{(4\pi)^2} m_b[\bar{s}\sigma^{\mu\nu} P_L b]F_{\mu\nu}\,,\quad \mathcal{O}^{\mu\prime}_{9(10)} = \frac{e^2}{(4\pi)^2}[\bar{s}\gamma^\mu P_R b][\bar{\mu}\gamma_\mu(\gamma_5)\mu] \ .
\end{split}\end{equation}
The four-quark operators $\mathcal{O}_{1,\dots,6}^{(\prime)}$, instead enter through penguin diagrams (Figure~\ref{fig:currcurr}) and the chromomagnetic operator $\mathcal{O}_8^{(\prime)}$ enters via its mixing into $\mathcal{O}_7^{(\prime)}$. The effect of the former is often absorbed into ``effective'' WCs,\footnote{We do not consider $C_{7,9}^{{\rm eff}\prime}$ since contributions to $C_{1,\dots,6}^\prime$ in NP scenarios are either heavily constrained or very small~\cite{Altmannshofer:2008dz}.}
\begin{equation}
    C_7^{\rm eff} = C_7 - \frac{1}{3}C_3 - \frac{4}{9} C_4 - \frac{20}{3}C_5 - \frac{80}{9}C_6 + \Delta C_7^{\rm eff}\,, \quad C_9^{\rm eff} = C_9 + Y(q^2) + \Delta C_9^{\rm eff}\ ,
\end{equation}
where
\begin{equation}\begin{split}
    Y(q^2) =& \frac{4}{3}C_3 + \frac{64}{9}C_5 + \frac{64}{27}C_6 - \frac{1}{2}h(q^2,0)\Bigg(C_3 + \frac{4}{3}C_4 + 16 C_5 + \frac{64}{3}C_6\Bigg) \\[0.5em]
    &+ h(q^2,m_c)\Bigg(\frac{4}{3}C_1 + C_2 + 6 C_3 + 60 C_5\Bigg) \\[0.5em]
    &- \frac{1}{2}h(q^2, m_b)\Bigg(7 C_3 + \frac{4}{3}C_4 + 76 C_5 + \frac{64}{3} C_6\Bigg)\,, \\[0.5em]
    h\big(q^2, m\big) =& -\frac{4}{9}\Bigg(\log\Big(\frac{m^2}{\mu^2}\Big)- \frac{2}{3} - x\Bigg)\\[0.5em] & - \frac{4}{9}\Big(2 + x\Big)\times\left\{
    \begin{split}
        \sqrt{x - 1}\arctan\frac{1}{\sqrt{x - 1}} \hspace{1.8cm} &\,x > 1 \\
        \sqrt{1 - x}\Big(\log\frac{1 + \sqrt{1 - x}}{\sqrt{x}} - \frac{i\pi}{2}\Big) \quad &\,x\leq 1
    \end{split}
    \right.\,\,\,,
\end{split}\end{equation}
with $x = 4m^2/q^2$. The $O(\alpha_s)$ QCD contributions to non-local matrix elements arising from insertions of current-current and chromomagnetic operators, given in Refs.~\cite{Ghinculov:2003qd,Greub:2008cy}, are contained in $\Delta C_{7,9}^{\rm eff}$.

\subsection{Formalism of Differential Measurement~\label{sec:fdmeas}}
The spin-summed differential decay of the $B_s^0\to\phi(\to K^+  K^-) \mu^+\mu^-$ can be written as \cite{Altmannshofer:2008dz}:
\begin{align}
    \frac{d^4\Gamma_{B_s^0\to\phi\mu^+\mu^-}}{dq^2 \,d\cos{\theta_K}\,d\cos{\theta_\ell}\,d\phi} &= \sum_i J_i(q^2) f_i(\theta_K, \theta_\ell, \phi)\nonumber\\
    &= \frac{9}{32\pi} ( J_{1s}\sin^2\theta_K+J_{1c}\cos^2\theta_K+J_{2s}\sin^2\theta_K\cos{2\theta_\ell}\nonumber\\
    &\qquad\quad+J_{2c}\cos^2\theta_K\cos{2\theta_\ell} +J_{3}\sin^2\theta_K\sin^2{\theta_\ell}\cos{2\phi}\nonumber\\
    &\qquad\quad+J_{4}\sin 2\theta_K\sin{2\theta_\ell}\cos{\phi}+J_{5}\sin 2\theta_K\sin{\theta_\ell}\cos{\phi}\nonumber\\
    &\qquad\quad+J_{6s}\sin^2\theta_K\cos{\theta_\ell}+J_{6c}\cos^2\theta_K\cos{\theta_\ell}\nonumber\\
    &\qquad\quad+J_{7}\sin 2\theta_K\sin{\theta_\ell}\sin{\phi}+J_{8}\sin 2\theta_K\sin{2\theta_\ell}\sin{\phi}\nonumber\\
    &\qquad\quad+J_{9}\sin^2\theta_K\sin^2{\theta_\ell}\sin{2\phi}) \ ,
    \label{eq:diffDR}
\end{align}
which is expressed in terms of the angular coefficients $J_i(q^2)$ and the angular variables $\theta_K$, $\theta_\ell$ and $\phi$. In our study, we do not consider specific angular observables, instead integrating over the angular variables to obtain
\begin{align}
    \frac{d\Gamma_{B_s^0\to\phi\mu^+\mu^-}}{dq^2} = \frac{3}{4} \left( 2 J_{1s}+J_{1c}\right)-\frac{1}{4}\left(2J_{2s}+J_{2c}\right) \ , 
    \label{eq:fdr}
\end{align}
where the relevant angular coefficients are given by: 
\begin{equation}\begin{split}
J_{1s} &= \frac{(2+\beta_\mu^2)}{4} \left( |A_\perp^L|^2 + |A_\parallel^L|^2 + |A_\perp^R|^2 + |A_\parallel^R|^2 \right) + \frac{4m_\mu^2}{q^2}\Re \left(A_\perp^L A_\perp^{R\ast}+A_\parallel^L A_\parallel^{R\ast} \right) \ ,\\
J_{1c} &= |A_0^L|^2+|A_0^R|^2+\frac{4m_\mu^2}{q^2} \left[|A_t|^2+2\Re\left(A_0^LA_0^{R\ast}\right) \right] \ ,\\
J_{2s} &= \frac{\beta_\mu^2}{4}\left( |A_\perp^L|^2+ |A_\parallel^L|^2+ |A_\perp^R|^2+ |A_\parallel^R|^2\right) \  ,\\
J_{2c} &= -\beta_\mu^2\left(|A_0^L|^2 +|A_0^R|^2 \right)\ , 
\label{eq:J}
\end{split}\end{equation}
with $\beta_\mu = \sqrt{1-4m_\mu^2/q^2}$. Here, $A_X$ represent transversity amplitudes of the $B_s^0\to\phi(\to K^+K^-)\mu^+\mu^-$ decay, which can be found in Appendix~\ref{app:Amp}.

The spin-summed differential decay rate of $\bar{B}_s^0\to\phi(\to K^+  K^-) \mu^+\mu^-$ can be obtained by replacing $J_i$ with $\widetilde{J}_i=\zeta_i \bar{J}_i$ in Eq.~\eqref{eq:diffDR}, where $\zeta_i=1$ for the final state considered and hence $\widetilde{J}_i = \bar{J}_i$ for $i=1s,2c,2s,2c$ in the scope of this study. To obtain $\bar{J}_i$ from $J_i$, one makes the replacement $A_X \to \bar{A}_X$ by changing the sign of all the weak phases. More details can be found in~\cite{Descotes-Genon:2015hea}. Hence, we can write:
\begin{align}
    \frac{d\Gamma_{\bar{B}_s^0\to\phi\mu^+\mu^-}}{dq^2} &= \frac{3}{4} ( 2 \widetilde{J}_{1s}+\widetilde{J}_{1c})-\frac{1}{4}(2\widetilde{J}_{2s}+\widetilde{J}_{2c})
    \ ,
    \label{eq:timeJanti}
\end{align}
which is the same as $d\Gamma_{B_s^0\to\phi\mu^+\mu^-}/dq^2$ in the SM. This is because the angular coefficients we are concerned with, shown in Eq.~\eqref{eq:J}, consist only of the magnitudes of $A_X$ and the real part of products of $A_X$ and $A_Y^\ast$. Therefore, changing the sign of the weak phase in $A_X$ would result in $\bar{J}_i=J_i$. This also suggests that there is no $C\!P$ asymmetry in this decay, as
\begin{align}
    \langle A_{C\!P}\rangle 
    = \int dq^2\, \left(\frac{d\Gamma_{\bar{B}_s^0\to\phi\mu^+\mu^-}}{dq^2}-\frac{d\Gamma_{{B}_s^0\to\phi\mu^+\mu^-}}{dq^2} \right)= 0\ .
\end{align}
When also considering the time-dependence of the differential decay rate, the transversity amplitudes must be modified accordingly to:
\begin{align}
    A_X(t)=g_{+}(t) A_X + \frac{q}{p} g_-(t)\widetilde{A}_X \ ,\\
    \widetilde{A}_X(t)=\frac{p}{q} g_{-}(t) A_X + g_+(t)\widetilde{A}_X \ ,
\end{align}
where $\widetilde{A}_X$ represent transversity amplitudes of the $\bar{B}_s^0\to\phi(\to K^+  K^-) \mu^+\mu^-$ decay and $g_\pm(t)$ are time-dependent coefficients arising from $B_s^0-\bar B_s^0$ mixing. Then, the angular coefficients are modified by changing $A_X\to A_X(t)$. This leads to
\begin{align}
    J_i(t) + \widetilde{J}_i(t) &= e^{-\Gamma_s t}\left[(J_i+ \widetilde{J}_i)\cosh{\frac{1}{2}\Delta\Gamma_s t}-h_i \sinh{\frac{1}{2}\Delta\Gamma_s t}\right] \ ,\text{ and}
    \label{eq:timeJ1}\\
    J_i(t) - \widetilde{J}_i(t) &= e^{-\Gamma_s t}\left[(J_i- \widetilde{J}_i)\cos{\Delta m_s t}-s_i \sin{\Delta m_s t}\right] \ .
    \label{eq:timeJ2}
\end{align}
Then the time-dependent $A_{C\!P}(t)$ can be expressed by using Eqs.~\eqref{eq:timeJ1} and~\eqref{eq:timeJ2} in Eqs.~\eqref{eq:fdr} and~\eqref{eq:timeJanti}, giving
\begin{equation}\label{eq:JACPtime}
    A_{C\!P}(t) = \frac{\int dq^2\sum_i\kappa_i\Big[\big(J_i(q^2) - \tilde{J}_i(q^2)\big)\cos\big(\Delta m_{B_s} t\big) - s_i \sin\big(\Delta m_{B_s}t\big)\Big]}{\int dq^2\sum_i\kappa_i\Big[\big(J_i(q^2) + \tilde{J}_i(q^2)\big)\cosh\big(\Delta \Gamma_{B_s} t/2\big) - h_i \sinh\big(\Delta \Gamma_{B_s}t/2\big)\Big]}\,,
\end{equation}
where $i=1s,1c,2s,$ and $2c$; the coefficients are given by
\begin{equation}
    \kappa_{1s} = \frac{3}{2}, \qquad \kappa_{1c} = \frac{3}{4}, \qquad \kappa_{2s} = -\frac{1}{2}, \qquad \kappa_{2c} = -\frac{1}{4}\,,
\end{equation}
and the explicit expressions for $s_i$ and $h_i$ are given in Appendix~\ref{app:Amp}.
We can use Eqs.~\eqref{eq:fdr} and~\eqref{eq:timeJanti} to give
\begin{equation}
    \int dq^2\sum_i \kappa_i\big(J_i(q^2) + \tilde{J}_i(q^2)\big) = \Gamma(B^0_s\to\phi\mu^+\mu^-) + \Gamma(\bar B^0_s\to\phi\mu^+\mu^-)\,.
\end{equation}
Then, the coefficients in Eq.~\eqref{eq:acpTime} are given by
\begin{equation}\begin{split}\label{eq:observables}
    &C_{\phi\mu\mu} = \frac{\int dq^2\sum_i\kappa_i\big(J_i(q^2) - \tilde{J}_i(q^2)\big)}{\Gamma + \bar \Gamma}, \quad S_{\phi\mu\mu} = -\frac{\int dq^2\sum_i\kappa_i\,s_i}{\Gamma + \bar\Gamma}, \\[0.5em]
    &\hspace{2.5cm} \text{and }\hspace{0.5cm} D_{\phi\mu\mu} = -\frac{\int dq^2\sum_i\kappa_i\,h_i}{\Gamma + \bar\Gamma}\,,
\end{split}\end{equation}
where we define $\Gamma = \Gamma_{B^0_s\to\phi\mu^+\mu^-}$ and $\bar\Gamma = \Gamma_{\bar B^0_s\to\phi\mu^+\mu^-}$. Note that, in the following, we consider complex NP WCs, in which case $J_i(q^2)\neq \tilde{J}_i(q^2)$.

The numerators and denominators of the observables in Eq.~\eqref{eq:observables} are each polynomial functions of the WCs. Since the coefficients of these -- at most quadratic -- polynomials are independent of the BSM WCs, it is simplest for our analysis to extract these coefficients from the theory prediction. We therefore will be interested in the following quantities:
\begin{equation}\label{eq:tildCoefs}
    \begin{split}
        &\Gamma^R = \int_R dq^2\sum_i \kappa_i J_i(q^2), \quad \bar\Gamma^R = \int_R dq^2\sum_i \kappa_i\tilde{J}_i(q^2), \quad \tilde C_{\phi\mu\mu}^R = \Gamma^R - \bar\Gamma^R, \\[0.5em]
        &\hspace{2cm}\tilde{S}_{\phi\mu\mu}^R = -\int_Rdq^2\sum_i \kappa_i s_i, \quad \tilde{D}_{\phi\mu\mu}^R = -\int_Rdq^2\sum_i \kappa_i h_i\,,
    \end{split}
\end{equation}
from which, it is straightforward to recover the observables in Eq.~\eqref{eq:observables}.

\subsection{Constraints on Wilson Coefficients}\label{sec:npparam}
We can now use the WET framework to express the observables introduced in Section~\ref{sec:results} in terms of the WCs. Using the expected sensitivities at the FCC-$ee$, we can then place constraints on these WCs. The sensitivities of operators in the WET are different between the high- and low-$q^2$ regions. We therefore evaluate each obervable in these regions separately, as well as performing an analysis of the two regions combined (discussed below). Additionally, we separate the NP WCs into real and imaginary parts as $\delta C_x = \delta C_{xR} + i\delta C_{xI}$.

The flavor-specific decay rate in the respective $q^2$ region, $R$, is then given as
\begin{equation}\begin{split}\label{eq:NPBrs}
    &\Gamma^{R}\times 10^{19} = \Gamma^{R}_{\rm SM}\times 10^{19} + \sum_k b^{R}_k \delta C_k + \sum_{k\ell}B^{R}_{k\ell}\delta C_k\delta C_\ell\,,
\end{split}\end{equation}
where $k,\ell\in\{7R,7I,9R,9I,10R,10I\}$. Similar expressions for $\bar\Gamma$ are found by taking $b_{xI}\to- b_{xI}$, $B_{xI,yR} \to - B_{xI,yR}$, and $B_{xR,yI}\to - B_{xR,yI}$ in Eq.~\eqref{eq:NPBrs}, where $x,y\in\{7,9,10\}$. We note that $\Gamma^R = \bar\Gamma^R$ in the SM.

In order to evaluate constraints on WCs using the measured branching ratio at the FCC-$ee$, mixing effects must also be accounted for. To this end, we use the time-averaged decay rate,
\begin{equation}
    \left<\Gamma_{B_s^0\to \phi\mu^+\mu^-}\right> = \frac{1}{2(1 - y^2)}\int_R dq^2\sum_i \kappa_i\Big(J_i(q^2) + \bar J_i(q^2) - y h_i(q^2)\Big)\,,
\end{equation}
where $y = \Delta \Gamma/2\Gamma = 0.06$ and $\bar J_i = \tilde J_i$ for all the relevant values of $i$~\cite{Descotes-Genon:2015hea,HFLAV:2014fzu}. In particular, we use the ``hadronic'' form of the integrated observables given in Ref.~\cite{Descotes-Genon:2015hea} due to the fact that the $B_s$ and $\bar B_s$ are produced incoherently at the $Z$-pole~\cite{Wang:2022nrm}. We again expand the time-averaged decay rate as a quadratic polynomial in BSM WCs,
\begin{equation}\begin{split}\label{eq:tavNPBrs}
    &\left<\Gamma\right>^R\times 10^{19} = \left<\Gamma\right>^{R}_{\rm SM}\times 10^{19} + \sum_k \left<b\right>^{R}_k \delta C_k + \sum_{k\ell}\left<B\right>^{R}_{k\ell}\delta C_k\delta C_\ell\,.
\end{split}\end{equation}

The coefficients in Eq.~\eqref{eq:tildCoefs} can be written in an analogous way as
\begin{equation}\begin{split}\label{eq:coefExpand}
    &\tilde{C}^{R}_{\phi\mu\mu}\times 10^{19} = \sum_k\,\gamma_k^R\,\delta C_k + \sum_{k\ell}G_{k\ell}^R\,\delta C_k\delta C_\ell\,, \\[0.5em]
    &\tilde{S}^{R}_{\phi\mu\mu}\times 10^{19} = \sum_k\,\sigma_k^R\,\delta C_k + \sum_{k\ell}\Sigma_{k\ell}^R\,\delta C_k\delta C_\ell\,, \\[0.5em]
    &\tilde{D}^R_{\phi\mu\mu}\times 10^{19} = \tilde{D}^R_{\phi\mu\mu,{\rm SM}}\times 10^{19} + \sum_k\,\delta_k^R\,\delta C_k + \sum_{k\ell}\Delta_{k\ell}^R\,\delta C_k\delta C_\ell\,.
\end{split}\end{equation}

To avoid the theoretical challenges associated with the intermediate-$q^2$ region--where narrow charmonium resonances complicate calculations--while still retaining a high event yield, fits are performed over the compound range  $R' = [0.1\text{ GeV}^2, 8.0\text{ GeV}^2] \cup [15.0\text{ GeV}^2, q^2_{\text{max}}]$. The SM values of the observables of interest in this range are
\begin{equation}\begin{split}
    &\Gamma^{R'}_{\text{SM}} = \bar \Gamma^{R'}_{\text{SM}} = 3.30(16)\times 10^{-19} \text{ GeV}\,, \\[0.5em]
    &\left<\Gamma\right>^{R'}_{\text{SM}} = 3.16(30)\times 10^{-19} \text{ GeV}\,, \\[0.5em]
    &\tilde D^{R'}_{\phi\mu\mu,\text{SM}} = - 4.96(56)\times 10^{-19} \text{ GeV}\,,
\end{split}\end{equation}
and $\tilde{S}^{R'}_{\text{SM}} = \tilde{C}^{R'}_{\text{SM}} = 0$.\footnote{Standard model values have been checked using the open-source codes \texttt{flavio}~\cite{Straub:2018kue} and \texttt{EOS}~\cite{vanDyk:2021sup}. Percent-level discrepancies arise in comparisons with \texttt{flavio} primarily due to different inputs and the fact that we neglect two-loop up-quark contributions to the matrix elements which are included in \texttt{flavio}. Larger differences are found in comparisons with \texttt{EOS}, particularly in the low-$q^2$ region, which we mostly attribute to the fact that \texttt{EOS} includes contributions from non-local form factors~\cite{Gubernari:2022hxn}. As these contributions cannot be extended to the high-$q^2$ region, we do not include these effects.} Expansion coefficients for BSM WCs in this range are given in Appendix~\ref{app:coefs}. Furthermore, we also include the more standard ranges of $q^2\in [1.1, 6.0]$ GeV$^2$ and $q^2\in [15.0\text{ GeV}^2, q^2_{\text{max}}]$, denoted as the low- and high-$q^2$ regions, respectively, in our fit using the time-averaged branching ratio. This is primarily due to the fact that the branching ratio is sensitive to different values of BSM WCs in these two ranges and this observable is not statistically limited by time-binning. The SM predictions of the decay rate in these ranges are given by
\begin{equation}
    \left<\Gamma\right>^{\text{Low}}_{\text{SM}} = 1.16(13)\times 10^{-19}\text{ GeV}\,, \qquad
    \left<\Gamma\right>^{\text{High}}_{\text{SM}} = 1.00(11)\times 10^{-19}\text{ GeV}\,.
\end{equation}

\begin{figure*}[h!]
    \centering
    \begin{subfigure}[t]{1\textwidth}
        \centering
        \includegraphics[width=1\textwidth]{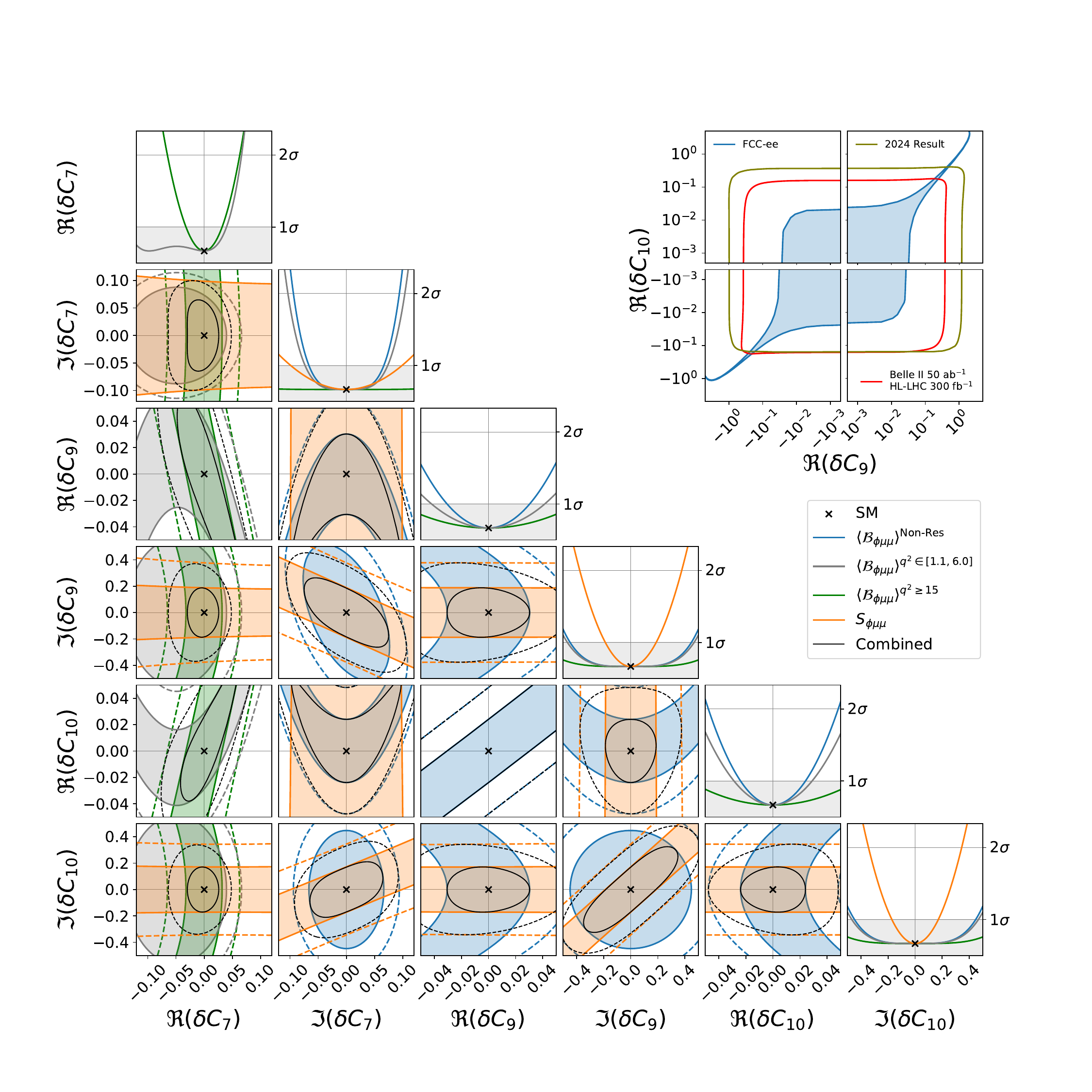}
    \end{subfigure}%
    \caption{The 2D constraints and the 1D likelihood of the corresponding WCs. The shaded regions represent $1\sigma$ contours and the dashed lines represent $2\sigma$ contours. The black color represents the combined contours of the measurements. Upper right panel shows the comparison of the $1\sigma$ constraints of future projection from Belle~II and HL-LHC combined~\cite{Huber:2024rbw}, and our FCC-$ee$ projection. This shows that the FCC-$ee$ projection can reach $\mathcal{O}(10)$ improvement with respected to the pre-FCC experiments projection. }
    \label{fig:WC}
\end{figure*}

The best 2D constraints and 1D likelihoods on the WCs, derived from the observables discussed, are presented in Figure~\ref{fig:WC}. In these plots, solid (dashed) lines indicate the $1\sigma$ ($2\sigma$) confidence regions. The blue and orange contours correspond to the branching ratio and the $S_{\phi\mu\mu}$ measurements in the $R'$ range introduced above, respectively. The grey and green contours represent the branching ratio measurements in the low- and high-$q^2$ regions, respectively, which exhibit different sensitivities, especially to the $\Re{(\delta C_7)}$ operator (see Figure~\ref{fig:WC_q2} in Appendix~\ref{app:WCconst}). The black contours represent the combined constraints. The 2D contour plots illustrate that branching ratio measurements are generally more sensitive to the real parts of the WCs, whereas the $C\!P$-violating observables ($S_{\phi\mu\mu}$ here) are primarily sensitive to the imaginary components. Therefore, these two types of observables serve as complementary probes. For example, in the $\Re{(\delta C_9)}$–$\Im{(\delta C_9)}$ plane, there exists a degeneracy in constraining $\Im{(\delta C_9)}$ using only branching ratio measurements. This degeneracy can be resolved by including the $S_{\phi\mu\mu}$ measurement.

The contour plot shown in the upper-right corner of Figure~\ref{fig:WC} provides a direct comparison between the FCC-$ee$ projection in the $\Re{(\delta C_9)}$–$\Re{(\delta C_{10})}$ plane and the projection from pre-FCC experiments~\cite{Huber:2024rbw}. The results clearly indicate that the FCC-$ee$ sensitivity could improve constraints on these WCs by approximately an order of magnitude.

\begin{figure*}[h!]
    \centering
    \begin{subfigure}[t]{1\textwidth}
        \centering
        \includegraphics[width=1\textwidth]{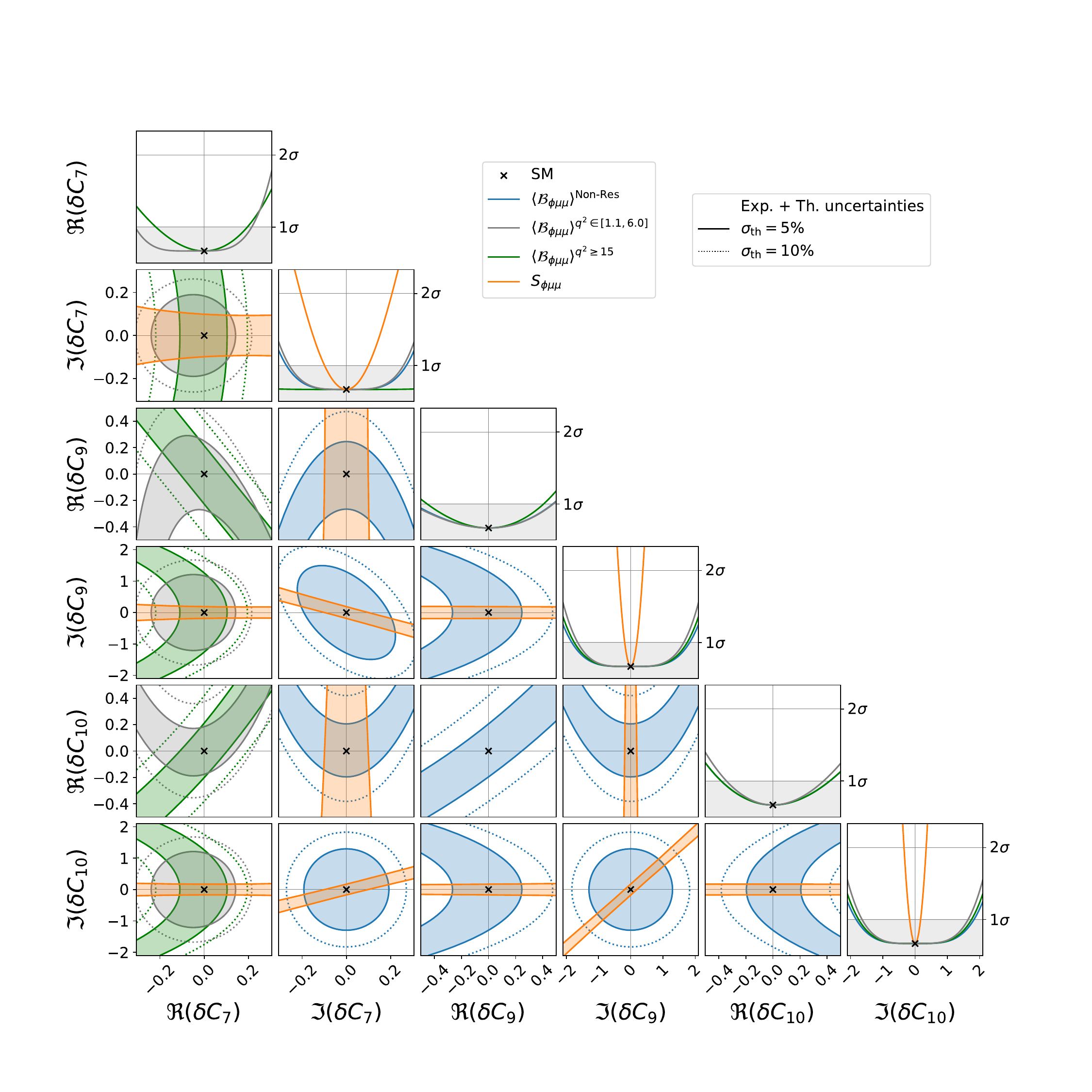}
    \end{subfigure}%
    \caption{The 2D constraints and the 1D likelihood of the corresponding WCs. The shaded regions represent $1\sigma$ contours of FCC experimental and the projected future theoretical uncertainty of $5\%$, and dotted represent the FCC experimental and the current theoretical uncertainty of $10\%$.}
    \label{fig:WC_th}
\end{figure*}

In addition to projections based solely on statistical uncertainties, we also consider the impact of theoretical uncertainties, as shown in Figure~\ref{fig:WC_th}. Related FCC-$ee$ projections for similar decay modes are discussed in Ref.~\cite{Bordone:2025cde}. We examine two theoretical scenarios: one with a theoretical uncertainty of $\sigma_{\text{th}} = 10\%$, which reflects the current level of uncertainty (solid lines), and another with an improved uncertainty of $\sigma_{\text{th}} = 5\%$, projected for the time of the {FCC-$ee$} era (dotted lines). It is important to note that theoretical uncertainties largely cancel in $C\!P$-violating observables such as $S_{\phi\mu\mu}$, meaning their constraints are unaffected by the size of $\sigma_{\text{th}}$. As a result, any observed deviation in such observables would serve as a more robust indication of potential NP, as opposed to branching ratio measurements, which may still be affected by theoretical ambiguities. This further underscores the value and significance of including $C\!P$ observables in precision studies at FCC-$ee$.

\begin{figure*}[h!]
    \centering
    \begin{subfigure}[t]{0.9\textwidth}
        \centering
        \includegraphics[width=0.95\textwidth]{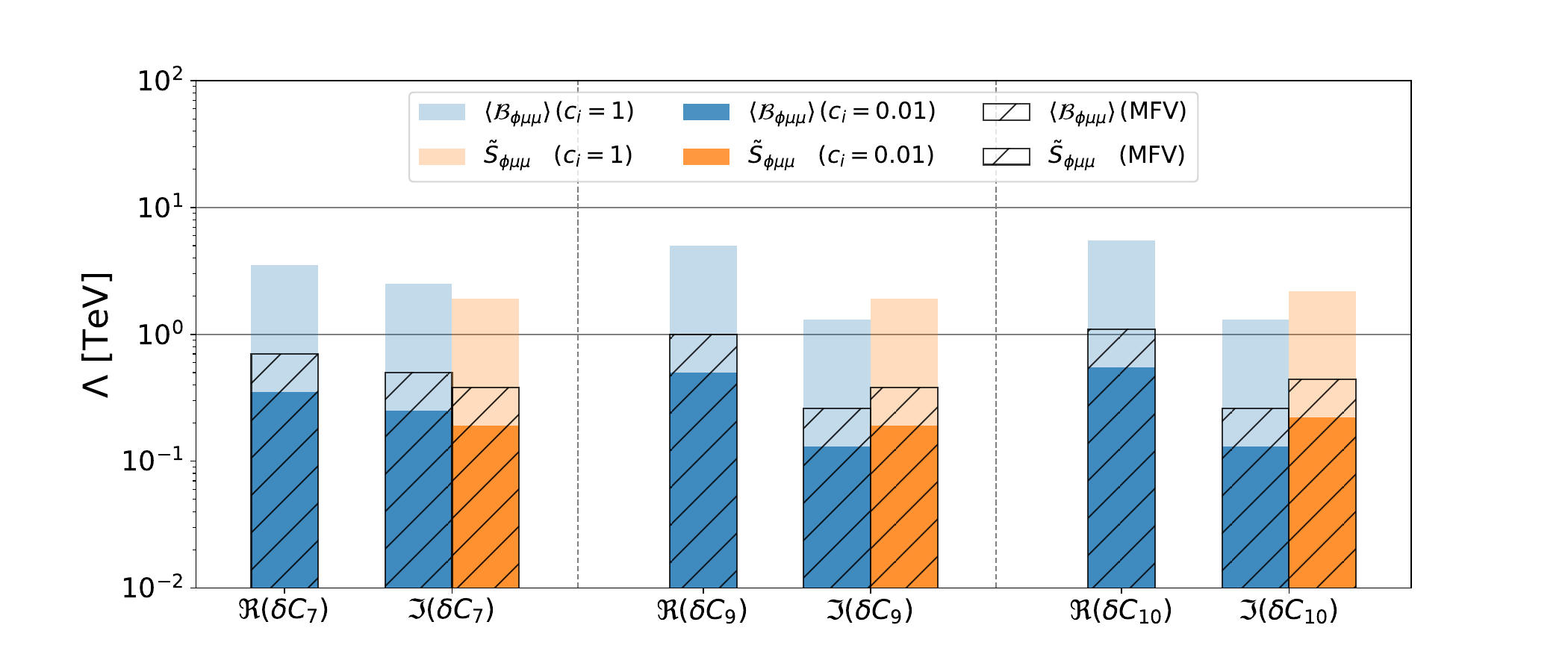}
    \end{subfigure}%
    \caption{The probe of the NP energy scale, $\Lambda$, of different WCs.  }
    \label{fig:Lambda}
\end{figure*}

Figure~\ref{fig:Lambda} illustrates the NP energy scales, $\Lambda$, that can be probed using the observables discussed above. The renormalization group running of the WCs is performed using the  Python package \texttt{wilson}~\cite{Aebischer:2018bkb}. In the plot, the blue and orange bars correspond to the energy probe derived from the branching ratio and $S_{\phi\mu\mu}$ measurements, respectively. Dark and light shades indicate two benchmark scenarios for the NP coupling strength, $c_i = 0.01$ and $c_i = 1$, while the hashed bars represent the case under the Minimal Flavor Violation (MFV) assumption. These results indicate that, assuming $c_i = 1$, the FCC-$ee$ can probe scales of $C\!P$-violating NP entering the $B^0_s\to\phi\mu^+\mu^-$ mode up to $\mathcal{O}(1$–$10)$~TeV.

More detailed plots of the constraints from all the observables discussed above can be found in Appendix~\ref{app:WCconst}.

\section{Conclusion}

This study explores the potential for performing time-dependent $C\!P$ measurements of rare $b \to s \mu^+ \mu^-$ decays at the $Z$-pole. To best demonstrate the feasibility of such measurements at future $Z$-factories, we conduct an in-depth simulation study focused on the $B_s^0 \to \phi \mu^+ \mu^-$ decay, utilizing \texttt{Pythia} for event generation and \texttt{Delphes} for detector simulation. A selection strategy based on sequential kinematic and particle identification criteria is employed, achieving a relative precision of 0.515\% on the branching ratio. This represents an improvement of approximately a factor of five over current LHCb results.

Beyond the branching ratio, we demonstrate the advantages offered by the FCC-$ee$’s excellent vertex resolution and flavor tagging capabilities for facilitating $C\!P$ measurements. These include: the untagged time-dependent decay rate, where the projected uncertainty on the observable $D_f$ is 0.103; the time-integrated $C\!P$ asymmetry, for which we estimate an uncertainty of $\sigma\left(\langle A_{C\!P} \rangle\right) = 9.21 \times 10^{-3}$, assuming a tagging power of $P_{\rm tag} = 0.3$; and the time-dependent $C\!P$ asymmetry, with projected uncertainties of 0.0210 and 0.0214 on the $C_f$ and $S_f$ parameters, respectively. These results highlight the strong potential of FCC-$ee$ for precision $C\!P$ measurements in rare $B$ decays.

In addition to the simulation study, we interpret the estimated experimental uncertainties within the framework of the WET. Our analysis demonstrates that the branching ratio measurement and the $C\!P$ observables serve as complementary probes: while the branching ratio is primarily sensitive to the real parts of the WCs, the $C\!P$ observables are more sensitive to the imaginary parts. Furthermore, our projections indicate that the FCC-$ee$ can achieve an improvement of approximately one order of magnitude in constraining the WCs compared to pre-FCC projections.

We also account for theoretical uncertainties in our projections. This analysis further underscores the importance of time-dependent $C\!P$ measurements, as these observables are largely insensitive to theoretical uncertainties, making them particularly robust indicators of NP. Therefore, any observed deviation in these $C\!P$ observables would strongly suggest the presence of BSM physics, rather than a limitation in theoretical understanding, as has been discussed extensively in the context of exclusive-mode branching ratios~\cite{Jager:2014rwa,Ciuchini:2020gvn,Ciuchini:2021smi,Ciuchini:2022wbq,Isidori:2024lng, Bordone:2024hui}.
Finally, we translate the projected constraints on the WCs into corresponding NP energy scales. Our results show that, under the assumption of a coupling strength $c_i = 1$, the combined constraints from branching ratio and $C\!P$ observables enable sensitivity to NP scales up to $\mathcal{O}(1$–$10)~\mathrm{TeV}$.

\acknowledgments
We would like to thank Gino Isidori for helpful discussions. We are also grateful to Armin Ilg for his valuable comments on the detector concept. Special thanks to Franco Grancagnolo and Margherita Primavera for their explanations about the simulation of the mis-ID rate. THK would like to thank Lingfeng Li for useful discussions. ZP would like to thank Peter Stangl, Aleks Smolkovi\v{c}, and Matthew Kirk for helpful discussions.

This work was supported by the Swiss National Science Foundation. Additional support was provided by the Swiss High Energy Physics initiative for the FCC (CHEF), with funding provided specifically by SERI and the University of Zurich. The work of J.A. is supported by the European Union’s Horizon 2020 research and innovation program under the Marie Skłodowska-Curie grant agreement No.~101145975-EFT-NLO.

\clearpage
\newpage
\appendix
\section{Mis-identified Background\label{app:misID}}
\begin{figure*}[h!]
    \centering
    \begin{subfigure}[t]{0.5\textwidth}
        \centering
        \includegraphics[width=0.95\textwidth]{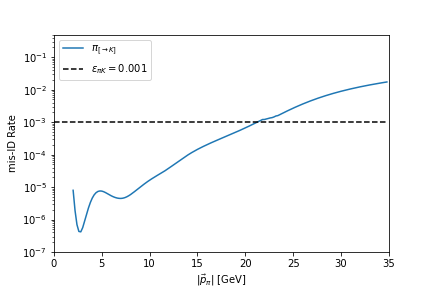}
    \end{subfigure}%
    ~ 
    \begin{subfigure}[t]{0.5\textwidth}
        \centering
        \includegraphics[width=0.95\textwidth]{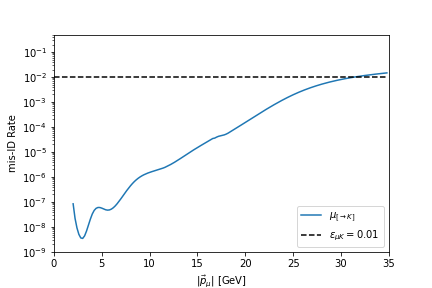}
    \end{subfigure}%
    
    \begin{subfigure}[t]{0.5\textwidth}
        \centering
        \includegraphics[width=0.95\textwidth]{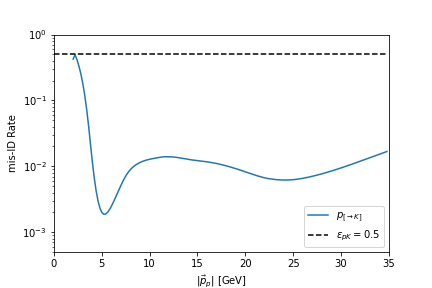}
    \end{subfigure}%
    ~ 
    \begin{subfigure}[t]{0.5\textwidth}
        \centering
        \includegraphics[width=0.95\textwidth]{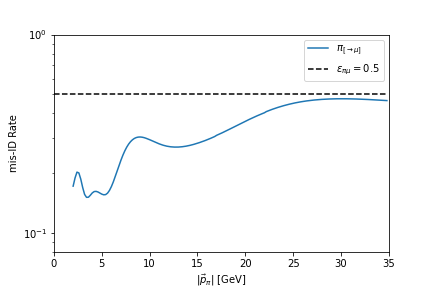}
    \end{subfigure}%

    \caption{The mis-ID rates as a function of momentum. The four panels show the mis-ID of different particles as discussed in the text. }
    \label{fig:misID}
\end{figure*}

We present a list of possible mis-ID backgrounds in Table~\ref{tab:misID}. In this context, we use the notation $i_{[\to j]}$ to indicate that particle $i$ is mis-identified as particle $j$. The table shows the expected events that will be produced in the FCC-$ee$, where $\epsilon_{ij}$ represents the probability of particle $i$ being mis-identified as particle $j$. The mis-ID rate is estimated for the cluster counting $dn/dx$ method in the IDEA concept with a drift chamber filled with $90\%$ He and $10\%$ iC$_4$H$_{10}$. The mis-ID rates ($\epsilon_{\pi_{[\to K]}}$, $\epsilon_{\mu_{[\to K]}}$, $\epsilon_{p_{[\to K]}}$ and $\epsilon_{\pi_{[\to \mu]}}$) as a function of the particle momentum are shown in Figure~\ref{fig:misID},\footnote{This part is done based on the work of Franco Grancagnolo and Margherita Primavera~\cite{defilippis2023cluster, primavera2024progress}. } with the dashed lines representing the benchmark values that are used to estimate the $s/b$ in Table~\ref{tab:misID}. The table also includes the expected signal-to-background ratio. For channels involving a muon as one of the mis-identified particles, we provide only a conservative estimate, excluding the information from the muon chamber. In fact, a lepton identification performance study has estimated $\epsilon_{\pi\mu}\sim 1\%$ when the energy of the pion is larger than $2$~GeV~\cite{Yu:2021pxc}.

Most of the entries have $s/b > \mathcal{O}(10^2)$, suggesting that they can be neglected in the leading-order approximation. However, some channels have $s/b < \mathcal{O}(10^2)$. These channels are still negligible in our analysis, as they will be highly suppressed by the selection requirements discussed in Section~\ref{sec:SimSel}. Specifically, those with resonance intermediate states ($\phi, J/\psi$, and $\psi$) will be suppressed by the $q^2$ selection requirement. For example, $B^0 \to K^{\ast0} (\to  K^+\pi^-_{[\to K^-]}) J/\psi(\to\mu^+\mu^-)$ will be masked because of $J/\psi\to\mu^+\mu^-$. And, those with mis-identified $K^\pm$ not having resonance at around $m_\phi$ will be suppressed by the $m_\phi$ selection requirement. For instance, $B^0 \to K^{\ast0} (\to  K^+\pi^-_{[\to \mu^-]}) J/\psi(\to\mu^+\mu^-_{[\to K^-]})$ will be highly suppressed because the $K^+K^-$ pair does not form a resonance around $m_\phi$. Therefore, the mis-ID backgrounds will not be simulated explicitly in this study.

\begin{table}[h!]
    \resizebox{\textwidth}{!}{\begin{tabular}{ccccc}
    \hline
        \hline
        Channel & Br$(H_b \to {\rm inter.})$ & Br$({\rm inter.}\to f)$ & Expected Events &  $S/B$\\
        \hline
        $B_s^0 \to \phi (\to K^+ K^-) \mu^+\mu^-$ & $8.38\times 10^{-7}$ & $4.91\times 10^{-1}$ & $7.82\times 10^{4}$ & --\\
        \hline
        
        $B^0 \to K^{\ast0} (\to  K^+ \pi^-_{[\to K^-]}) \mu^+\mu^-$ & $9.04\times 10^{-7}$ & $\sim 1$ & $\epsilon_{\pi K} \times 6.69\times 10^{5}$ & $\mathcal{O}(10)$\\
        $B^0 \to K^{\ast0} (\to  K^+\pi^-_{[\to K^-]}) J/\psi(\to\mu^+\mu^-)$ & $1.27\times 10^{-3}$ & $5.96\times 10^{-2}$ & $\epsilon_{\pi K} \times 5.60\times 10^{7}$ & $\mathcal{O}(1)$\\
        $B^0 \to K^{\ast0} (\to K^+\pi^-_{[\to K^-]}) \psi(\to X+ \mu^+\mu^-)$ & $5.90\times 10^{-4}$ & $4.47\times 10^{-2}$ & $\epsilon_{\pi K} \times 1.95\times 10^{7}$ & $\mathcal{O}(10)$\\
        $B^0 \to K^{\ast0} (\to K^+ \pi^-_{[\to K^-]}) \phi(\to\mu^+\mu^-)$ & $1.00\times 10^{-5}$ & $2.85\times 10^{-4}$ & $\epsilon_{\pi K}\times 2.11\times 10^{3}$ & $\mathcal{O}(10^5)$\\
        $\Lambda_b \to K^+ \bar{p}_{[\to K^-]} \mu^+\mu^-$ & $2.60\times 10^{-7}$ & -- & $\epsilon_{pK} \times 4.16\times 10^{4}$ & $\mathcal{O}(10)$\\
        \hline
        $B^0 \to K^{\ast0} (\to  K^+\pi^-_{[\to \mu^-]})  \mu^+\mu^-_{[\to K^-]}$ & $9.04\times 10^{-7}$ & $\sim 1$ & $\epsilon_{\pi\mu}\epsilon_{\mu K} \times 6.69\times 10^{5}$ & $>\mathcal{O}(10)$\\
        $B^0 \to K^{\ast0} (\to  K^+\pi^-_{[\to \mu^-]}) J/\psi(\to\mu^+\mu^-_{[\to K^-]})$ & $1.27\times 10^{-3}$ & $5.96\times 10^{-2}$ & $\epsilon_{\pi\mu}\epsilon_{\mu K} \times 5.60\times 10^{7}$ & $>\mathcal{O}(10^{-1})$\\
        $B^0 \to K^{\ast0} (\to K^+\pi^-_{[\to \mu^-]}) \psi(\to X+ \mu^+\mu^-_{[\to K^-]})$ & $5.90\times 10^{-4}$ & $4.47\times 10^{-2}$ & $\epsilon_{\pi\mu}\epsilon_{\mu K} \times 1.95\times 10^{7}$ & $>\mathcal{O}(1)$\\
        $B^0 \to K^{\ast0} (\to K^+ \pi^-_{[\to \mu^-]}) \phi(\to\mu^+\mu^-_{[\to K^-]})$ & $1.00\times 10^{-5}$ & $2.85\times 10^{-4}$ & $\epsilon_{\pi\mu}\epsilon_{\mu K}\times 2.11\times 10^{3}$ & $>\mathcal{O}(10^4)$\\
        $B_s^0 \to \phi (\to K^+ K^-_{[\mu^-]}) \mu^+_{[K^+]}\mu^-$ & $8.38\times 10^{-7}$ & $4.91\times 10^{-1}$ & $2 \epsilon_{\mu K}^2 \times 7.82\times 10^{4}$ & $>\mathcal{O}(10^4)$\\
        $B_s^0 \to \phi (\to K^+ K^-_{[\mu^-]}) J/\psi(\to \mu^+_{[K^+]}\mu^-$) & $1.04\times 10^{-3}$ & $2.93\times 10^{-2}$ & $2\epsilon_{\mu K}^2 \times 5.48\times 10^{6}$ & $>\mathcal{O}(10^2)$\\
        $B_s^0 \to \phi (\to K^+ K^-_{[\mu^-]}) \psi(\to \mu^+_{[K^+]}\mu^-$) & $5.30\times 10^{-4}$ & $2.19\times 10^{-2}$ & $2 \epsilon_{\mu K}^2 \times 2.09\times 10^{6}$ & $>\mathcal{O}(10^2)$\\
        $B_s^0 \to \phi (\to K^+ K^-_{[\mu^-]}) \phi(\to \mu^+_{[K^+]}\mu^-$) & $1.85\times 10^{-5}$ & $1.40\times 10^{-4}$ & $2 \epsilon_{\mu K}^2 \times 4.66\times 10^{2}$ & $>\mathcal{O}(10^6)$\\

        \hline
        \hline
        
\end{tabular}}
    \caption{Potential mis-ID backgrounds, where $i_{[\to j]}$ denotes that particle $i$ is mis-identified as particle $j$, with probability $\epsilon_{ij}$. We also denote ``${\rm inter.}$'' as the intermediate particle(s) in the decay chain.}
    \label{tab:misID}
\end{table}

\section{Fitting $m_{B_s^0}$ Distributions\label{app:mBs}}
We fit the signal $m_{B_s^0}$ distribution as a double sided crystal ball, where the probability density function is written as: 
\begin{empheq}[left={f(m; \mu, \sigma, \alpha_L, n_L, \alpha_R, n_R) \propto \empheqlbrace}]{equation}
  \begin{aligned}
      &A_L\, \left(B_L-\frac{m-\mu}{\sigma}\right)^{-n_L} & & \text{if } \frac{m-\mu}{\sigma}<-\alpha_L\\[1ex]
      &\exp{\left[-\frac{1}{2}\left(\frac{m-\mu}{\sigma}\right)^2\right]} & & \text{if } -\alpha_L<\frac{m-\mu}{\sigma}<\alpha_R\ ,\\[1ex]
      &A_R\, \left(B_R+\frac{m-\mu}{\sigma}\right)^{-n_R} & & \text{if } \frac{m-\mu}{\sigma}>\alpha_R\\[1ex]
  \end{aligned}
\end{empheq}
where $A_i=\left(n_i/|\alpha_i|\right)^{n_i}\,\exp{(-|\alpha_i|^2/2)}$ and $B_i=\left(n_i/|\alpha_i|\right)-|\alpha_i|$. The parameters to be fitted are $\mu,\ \sigma,\ \alpha_L,\ n_L,\ \alpha_R$ and $n_R$. 

As for both $Z \to b\bar{b}$ and $Z\to c\bar{c}$ backgrounds, we model them as an exponential decay function. The probability density function, with parameter $\lambda$ to be fitted, is given by: 
\begin{align}
    f(m;\lambda)\propto e^{-\lambda m} \ .
\end{align}

\begin{figure*}[h!]
    \centering
    \begin{subfigure}[t]{0.5\textwidth}
        \centering
        \includegraphics[width=0.95\textwidth]{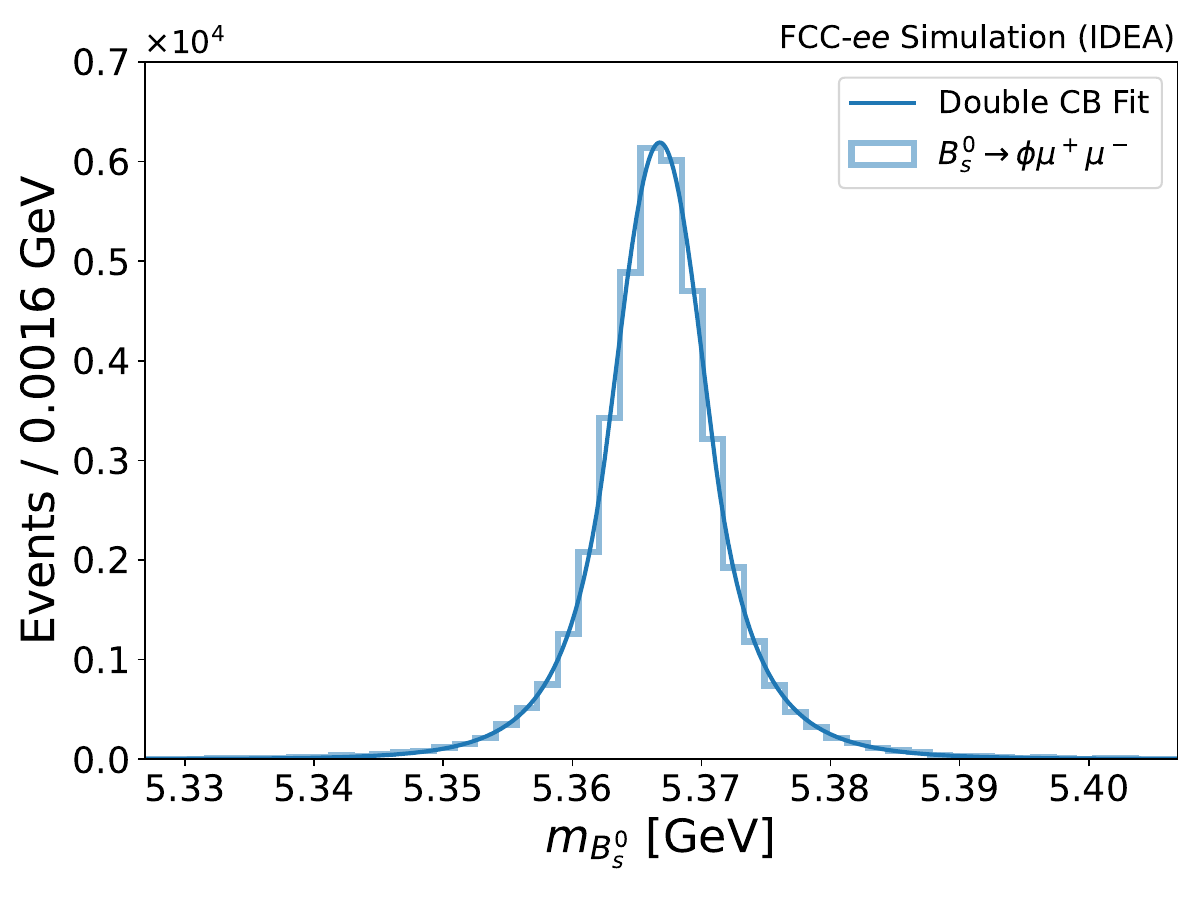}
    \end{subfigure}%
    
    \begin{subfigure}[t]{0.5\textwidth}
        \centering
        \includegraphics[width=0.95\textwidth]{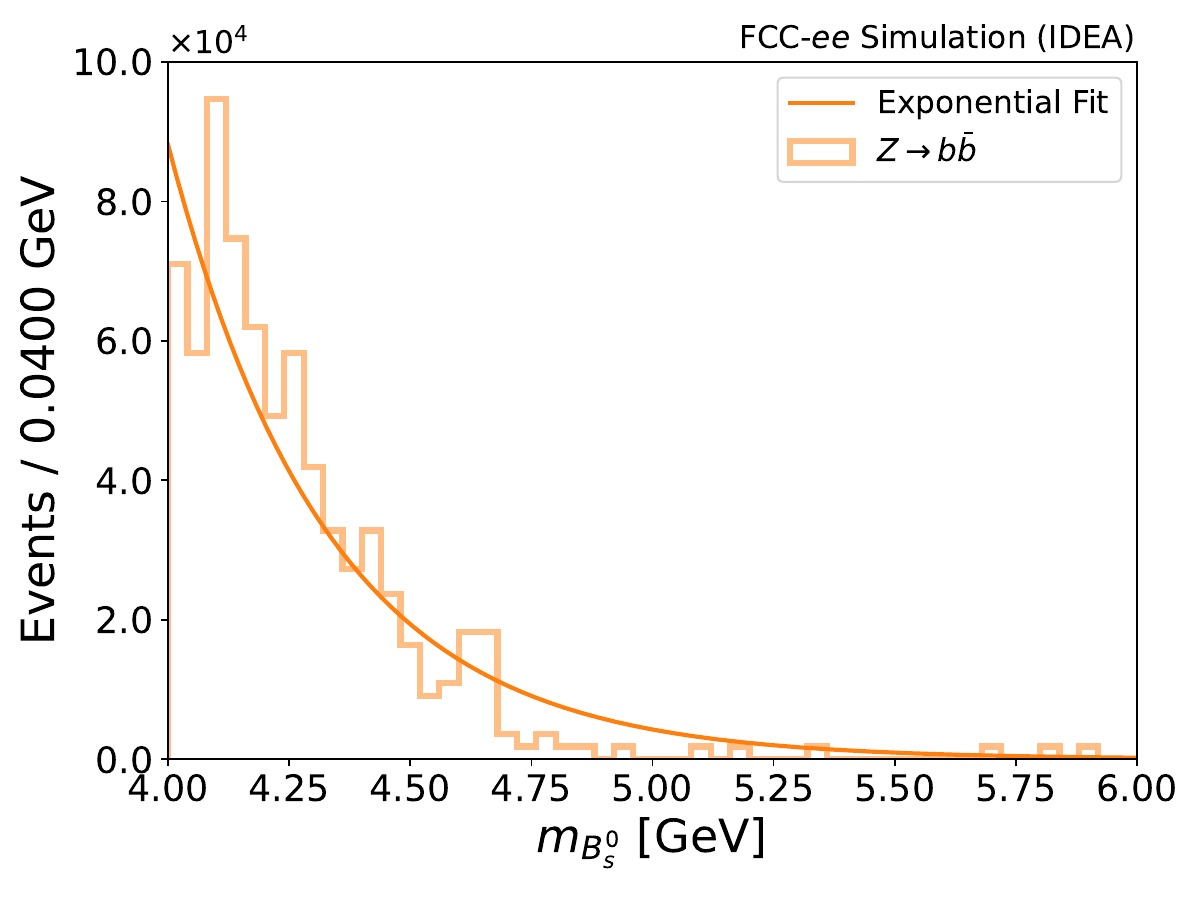}
    \end{subfigure}%
    ~ 
    \begin{subfigure}[t]{0.5\textwidth}
        \centering
        \includegraphics[width=0.95\textwidth]{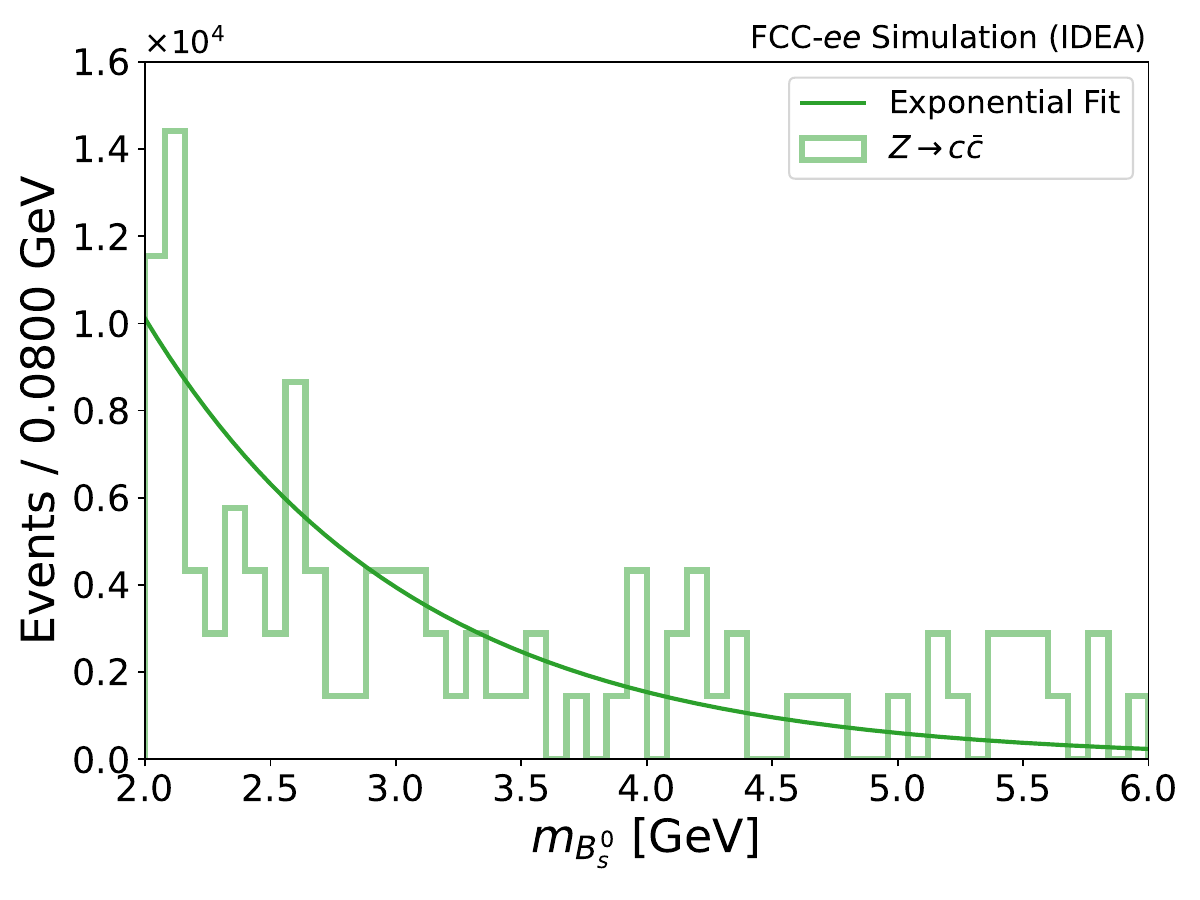}
    \end{subfigure}%
    
    \caption{The fitted distributions of reconstructed $B_s^0$ mass, from the MC samples after the $q^2$ selection requirement. }
    \label{fig:mBsFit}
\end{figure*}

The fitted results of the signal and backgrounds of are shown in Figure~\ref{fig:mBsFit}. 

\section{Statistical Uncertainty of $A_{C\!P}$ Measurement\label{app:acpUncert}}
We denote the asymmetry by $A=(n-\bar{n})/(n+\bar{n})$, where $n$ and $\bar{n}$ represent the number of $P^0\to f$ and $\bar{P}^0\to f$ decays, respectively. Assuming the numbers follow Poisson distributions with means $\lambda$ and $\bar{\lambda}$ ($n\sim {\rm Pois}(\lambda) $ and $\bar{n}\sim {\rm Pois}(\bar{\lambda}) $), the uncertainty of $A$ can be approximated by:
\begin{align}
    \sigma(A)^2 &\approx \sigma(n)^2 \left[\frac{\partial A}{\partial n}\bigg|_{(\lambda, \bar{\lambda})}\right]^2 +\sigma(\bar{n})^2 \left[\frac{\partial A}{\partial \bar{n}}\bigg|_{(\lambda, \bar{\lambda})}\right]^2 \nonumber\\
    &\approx \lambda \left[\frac{-2\lambda}{(\lambda+\bar{\lambda})^2}\right]^2 +\bar{\lambda} \left[\frac{2\bar{\lambda}}{(\lambda+\bar{\lambda})^2}\right]^2 \nonumber \\
    &\approx 4\frac{\lambda\bar{\lambda}}{(\lambda+\bar{\lambda})^3} \nonumber\\
    &\approx \frac{1-A^2}{N}\ .
\end{align}
Here, we define $N=\lambda+\bar{\lambda}$, then we can rewrite $\lambda=N(1+A)/2$, $\bar{\lambda}=N(1-A)/2$ and the expected $A=(\lambda+\bar{\lambda})/N$. 
The measured number of $P^0\to f$ and $\bar{P}^0\to f$ decays can be expressed as:
\begin{align}
    \lambda_{\rm meas} = (1-\omega)\lambda + \omega \bar{\lambda}\ , \quad \bar{\lambda}_{\rm meas} = \omega\lambda + (1-\omega) \bar{\lambda}\ , 
\end{align}
where $\omega$ is the wrong tagging fraction. This implies that the measured asymmetry can be expressed as 
\begin{align}
    \sigma(A) &= \frac{1}{1-2\omega}\sigma(A_{\rm meas}) \nonumber\\
    &\approx \frac{1}{\sqrt{\epsilon}(1-2\omega)}\sqrt{\frac{1-A_{\rm meas}^2}{N}}\nonumber\\
    &\approx \frac{1}{\sqrt{P_{\rm tag}}}\frac{1}{\sqrt{N}}\ ,
    \label{eq:acpApprox}
\end{align}
if we also take the tagging efficiency $\epsilon$ into account, which says that only $\epsilon N$ of the events are tagged. Using the definition of the tagging power $P_{\rm tag}\equiv \epsilon(1-2\omega)^2$, and the approximation that $A_{\rm meas}\ll 1$ in the channel that we are studying, we get the final form as shown in Eq.~\eqref{eq:acpApprox}. 

\section{Expressions of Amplitudes, $h_i$ and $s_i$ Functions\label{app:Amp}}
In this appendix, we present the analytic expressions of the tranversity amplitudes and $h_i$ and $s_i$ functions which enter the time-dependent observables in Eqs.~\eqref{eq:J} and~\eqref{eq:observables}.
They are given by (see also Appendix B in Ref.~\cite{Descotes-Genon:2022qce} and Appendix C of Ref.~\cite{Descotes-Genon:2015hea}):
\begin{align}
A_\perp^{L,R} &= N\sqrt{2\lambda}\left\{ \left(C_9 \mp C_{10}\right)\frac{V(q^2)}{m_{B_s^0}+m_{\phi}} +\frac{2m_b}{q^2}C_7 T_1(q^2)\right\}  \ ,\\
A_\parallel^{L,R} &= -N\sqrt{2}\left(m_{B_s^0}^2-m_{\phi}^2\right)\left\{   \left(C_9 \mp C_{10}\right)  \frac{A_1(q^2)}{m_{B_s^0}-m_{\phi}} +\frac{2m_b}{q^2}C_7 T_2(q^2)\right\} \ , \\
A_0^{L,R} &= -\frac{N}{2m_{\phi}\sqrt{q^2}}\Bigg\{ 2m_b C_7  \left[ \left(m_{B_s^0}^2+3m_{\phi}^2-q^2\right)T_2(q^2) - \frac{\lambda T_3(q^2)}{m_{B_s^0}^2-m_{\phi}^2}\right] \nonumber\\
&\quad + (C_9\mp C_{10}) \left [\left(m_{B_s^0}^2-m_{\phi}^2-q^2\right)\left(m_{B_s^0}+m_{\phi}\right)A_1(q^2) -\frac{\lambda A_2(q^2)}{m_{B_s^0}+m_{\phi}}\right ] \Bigg \}\ , \\
A_t &= 2N\sqrt{\frac{\lambda}{q^2}}C_{10} A_0(q^2) \ ,\\
    s_{1s}&=\frac{2+\beta_\mu^2}{2}\Im\left(\widetilde{A}^{L}_{\perp}A^{L*}_{\perp}+\widetilde{A}^{L}_{||}A^{L*}_{||}+\widetilde{A}^{R}_{\perp}A^{R*}_{\perp}+\widetilde{A}^{R}_{||}A^{R*}_{||}\right) \nonumber\\
    &\quad+\frac{4m_\mu^2}{q^2}\Im\left(\widetilde{A}^{L}_{\perp}A^{R*}_{\perp}+\widetilde{A}^{L}_{||}A^{R*}_{||}-A^{L}_{\perp}\widetilde{A}^{R*}_{\perp}-A^{L}_{||}\widetilde{A}^{R*}_{||}\right)\\
    s_{1c}&=2\Im\left(\widetilde{A}^{L}_0 A^{L*}_0+\widetilde{A}^{R}_0 A^{R*}_0\right)+\frac{8m_\mu^2}{q^2}{\Im}\left(\widetilde{A}_t  A^*_t+\widetilde{A}^{L}_0A^{R*}_0-A^{L}_0\widetilde{A}^{R*}_0\right)\\
    s_{2s}&=\frac{\beta_\mu^2}{2}{\Im}\left(\widetilde{A}^{L}_{\perp}A^{L*}_{\perp}+\widetilde{A}^{L}_{||}A^{L*}_{||}+\widetilde{A}^{R}_{\perp}A^{R*}_{\perp}+\widetilde{A}^{R}_{||}A^{R*}_{||}\right)\\
    s_{2c}&=-2\beta_\mu^2{\Im}\left(\widetilde{A}^{L}_0 A^{L*}_0+\widetilde{A}^{R}_0 A^{R*}_0\right)\\
    h_{1s}&=\frac{2+\beta_\mu^2}{2}\Re\left(\widetilde{A}^{L}_{\perp}A^{L*}_{\perp}+\widetilde{A}^{L}_{||}A^{L*}_{||}+\widetilde{A}^{R}_{\perp}A^{R*}_{\perp}+\widetilde{A}^{R}_{||}A^{R*}_{||}\right) \nonumber\\
    &\quad+\frac{4m_\mu^2}{q^2}\Re\left(\widetilde{A}^{L}_{\perp}A^{R*}_{\perp}+\widetilde{A}^{L}_{||}A^{R*}_{||}+A^{L}_{\perp}\widetilde{A}^{R*}_{\perp}+A^{L}_{||}\widetilde{A}^{R*}_{||}\right)\\
    h_{1c}&=2\Re\left(\widetilde{A}^{L}_0 A^{L*}_0+\widetilde{A}^{R}_0 A^{R*}_0\right)+\frac{8m_\mu^2}{q^2}{\Re}\left(\widetilde{A}_t  A^*_t+\widetilde{A}^{L}_0A^{R*}_0+A^{L}_0\widetilde{A}^{R*}_0\right)\\
    h_{2s}&=\frac{\beta_\mu^2}{2}{\Re}\left(\widetilde{A}^{L}_{\perp}A^{L*}_{\perp}+\widetilde{A}^{L}_{||}A^{L*}_{||}+\widetilde{A}^{R}_{\perp}A^{R*}_{\perp}+\widetilde{A}^{R}_{||}A^{R*}_{||}\right)\\
    h_{2c}&=-2\beta_\mu^2{\Re}\left(\widetilde{A}^{L}_0 A^{L*}_0+\widetilde{A}^{R}_0 A^{R*}_0\right)\ ,
\end{align}
with
\begin{align}
N &= V_{tb}V_{ts}^{\ast}\sqrt{\frac{G_F^2 \alpha^2}{3\times 2^{10}\pi^5 m^3_{B_s^0}}\sqrt{\lambda} q^2 \beta_\mu} \  \text{ , and}\\
\lambda &= \left[m_{B_s^0}^2-\left(m_\phi-\sqrt{q^2}\right)^2 \right] \left[m_{B_s^0}^2-\left(m_\phi+\sqrt{q^2}\right)^2 \right]\ .
\end{align}
The functions $A_{0,1,2}(q^2)$, $V(q^2)$ and $T_{1,2,3}(q^2)$ are form factors of the $B_s^0\to \phi$ transition, which can be found in \cite{Horgan:2013hoa, Horgan:2015vla,Bharucha:2015bzk}.

\section{Explicit Values of Observable Coefficients}\label{app:coefs}

In this appendix, we give the explicit values of the coefficients introduced in Eqs.~\eqref{eq:NPBrs},~\eqref{eq:tavNPBrs}, and~\eqref{eq:coefExpand}. In the fits, theory errors are either neglected, or an overall theory error, $\sigma_{\text{th}}$, is used. We therefore only give the central values of coefficients in this Appendix. For a more sophisticated handling of theoretical errors in exclusive $b\to s\bar\ell\ell$ decays at future colliders, see e.g. Ref.~\cite{Bordone:2025cde}. All coefficients are in units of $10^{19}$ GeV.

The flavor-specific and time-averaged decay rate coefficients relevant to the $q^2$ region of interest are given by:\footnote{The $B$, $\Sigma$, $\Delta$, and $\Gamma$ matrices are symmetric, so we only report the upper-triangular parts.}
\begin{align}
    &b_{k} = \begin{pmatrix}
        -0.33 & \hspace{0.3cm} -0.544 & \hspace{0.3cm} 0.651 & \hspace{0.3cm} -0.002 & \hspace{0.3cm} -0.828 & \hspace{0.3cm} 0
    \end{pmatrix}\,, \\[0.5em]
    &B_{k\ell} = \begin{pmatrix}
        4.50 & \hspace{0.3cm} 0 & \hspace{0.3cm} 0.333 & \hspace{0.3cm} 0 & \hspace{0.3cm} 0 & \hspace{0.3cm} 0 \\[0.3em]
         & \hspace{0.3cm} 4.50 & \hspace{0.3cm} 0 & \hspace{0.3cm} 0.333 & \hspace{0.3cm} 0 & \hspace{0.3cm} 0 \\[0.3em]
          & \hspace{0.3cm}   & \hspace{0.3cm} 0.099 & \hspace{0.3cm} 0 & \hspace{0.3cm} 0 & \hspace{0.3cm} 0 \\[0.3em]
          & \hspace{0.3cm}   & \hspace{0.3cm}   & \hspace{0.3cm} 0.099 & \hspace{0.3cm} 0 & \hspace{0.3cm} 0 \\[0.3em]
          & \hspace{0.3cm}   & \hspace{0.3cm}   & \hspace{0.3cm}   & \hspace{0.3cm} 0.099 & \hspace{0.3cm} 0 \\[0.3em]
          & \hspace{0.3cm}   & \hspace{0.3cm}   & \hspace{0.3cm}   & \hspace{0.3cm}   & \hspace{0.3cm} 0.099
    \end{pmatrix}\, ,\\[0.5em]
    &\left<b\right>_{k} = \begin{pmatrix}
        -0.40 & \hspace{0.3cm} 0 & \hspace{0.3cm} 0.617 & \hspace{0.3cm} 0 & \hspace{0.3cm} -0.791 & \hspace{0.3cm} 0
    \end{pmatrix}\, \text{ , and } \\[0.5em]
    &\left<B\right>_{k\ell} = \begin{pmatrix}
        4.51 & \hspace{0.3cm} 0 & \hspace{0.3cm} 0.325 & \hspace{0.3cm} 0 & \hspace{0.3cm} 0 & \hspace{0.3cm} 0 \\[0.3em]
         & \hspace{0.3cm} 4.54 & \hspace{0.3cm} 0 & \hspace{0.3cm} 0.343 & \hspace{0.3cm} 0 & \hspace{0.3cm} 0 \\[0.3em]
          & \hspace{0.3cm}   & \hspace{0.3cm} 0.095 & \hspace{0.3cm} 0 & \hspace{0.3cm} 0 & \hspace{0.3cm} 0 \\[0.3em]
          & \hspace{0.3cm}   & \hspace{0.3cm}   & \hspace{0.3cm} 0.104 & \hspace{0.3cm} 0 & \hspace{0.3cm} 0 \\[0.3em]
          & \hspace{0.3cm}   & \hspace{0.3cm}   & \hspace{0.3cm}   & \hspace{0.3cm} 0.095 & \hspace{0.3cm} 0 \\[0.3em]
          & \hspace{0.3cm}   & \hspace{0.3cm}   & \hspace{0.3cm}   & \hspace{0.3cm}   & \hspace{0.3cm} 0.104
    \end{pmatrix}\, .
\end{align}
The coeffieients corresponding to the observables relevant to the time-dependent $C\!P$ asymmetry are
\begin{align}
    &\gamma_k = \begin{pmatrix}
        0 & \hspace{0.3cm} -1.09 & \hspace{0.3cm} 0 & \hspace{0.3cm} -0.004 & \hspace{0.3cm} 0 & \hspace{0.3cm} 0
    \end{pmatrix}\,, \\[0.5em]
    &\sigma_k = \begin{pmatrix}
        0 & \hspace{0.3cm} 2.27 & \hspace{0.3cm} 0 & \hspace{0.3cm} 1.18 & \hspace{0.3cm} 0 & \hspace{0.3cm} -1.29
    \end{pmatrix}\,, \\[0.5em]
    &\delta_k = \begin{pmatrix}
        - 2.27 & \hspace{0.3cm} 0 & \hspace{0.3cm} - 1.18 & \hspace{0.3cm} 0 & \hspace{0.3cm} 1.29 & \hspace{0.3cm} 0
    \end{pmatrix}\,, \\[0.5em]
    &\Sigma_{k\ell} = \begin{pmatrix}
        0 & \hspace{0.3cm} 0.542 & \hspace{0.3cm} 0 & \hspace{0.3cm} 0.295 & \hspace{0.3cm} 0 & \hspace{0.3cm} 0 \\[0.3em]
         & \hspace{0.3cm} 0 & \hspace{0.3cm} 0.295 & \hspace{0.3cm} 0 & \hspace{0.3cm} 0 & \hspace{0.3cm} 0 \\[0.3em]
          & \hspace{0.3cm}   & \hspace{0.3cm} 0 & \hspace{0.3cm} 0.154 & \hspace{0.3cm} 0 & \hspace{0.3cm} 0 \\[0.3em]
          & \hspace{0.3cm}   & \hspace{0.3cm}   & \hspace{0.3cm} 0 & \hspace{0.3cm} 0 & \hspace{0.3cm} 0 \\[0.3em]
          & \hspace{0.3cm}   & \hspace{0.3cm}   & \hspace{0.3cm}   & \hspace{0.3cm} 0 & \hspace{0.3cm} 0.154 \\[0.3em]
          & \hspace{0.3cm}   & \hspace{0.3cm}   & \hspace{0.3cm}   & \hspace{0.3cm}   & \hspace{0.3cm} 0
    \end{pmatrix}\,  \text{ , and }\\[0.5em]
    &\Delta_{k\ell} = \begin{pmatrix}
        -0.542 & \hspace{0.3cm} 0 & \hspace{0.3cm} -0.295 & \hspace{0.3cm} 0 & \hspace{0.3cm} 0 & \hspace{0.3cm} 0 \\[0.3em]
         & \hspace{0.3cm} 0.542 & \hspace{0.3cm} 0 & \hspace{0.3cm} 0.295 & \hspace{0.3cm} 0 & \hspace{0.3cm} 0 \\[0.3em]
          & \hspace{0.3cm}   & \hspace{0.3cm} -0.154 & \hspace{0.3cm} 0 & \hspace{0.3cm} 0 & \hspace{0.3cm} 0 \\[0.3em]
          & \hspace{0.3cm}   & \hspace{0.3cm}   & \hspace{0.3cm} 0.154 & \hspace{0.3cm} 0 & \hspace{0.3cm} 0 \\[0.3em]
          & \hspace{0.3cm}   & \hspace{0.3cm}   & \hspace{0.3cm}   & \hspace{0.3cm} -0.154 & \hspace{0.3cm} 0 \\[0.3em]
          & \hspace{0.3cm}   & \hspace{0.3cm}   & \hspace{0.3cm}   & \hspace{0.3cm}   & \hspace{0.3cm} 0.154
    \end{pmatrix}\, .
\end{align}
Note that the $\Gamma_{k\ell}$ coefficients are not included, as they vanish.

The corresponding coefficients for the time-averaged decay rate in the low- and high-$q^2$ regions are given by
\begin{align}
    \left<b\right>_k^{\text{Low}} =& \begin{pmatrix}
        0.169 & \hspace{0.3cm} 0 & \hspace{0.3cm} 0.225 & \hspace{0.3cm} 0 & \hspace{0.3cm} - 0.326 & \hspace{0.3cm} 0
    \end{pmatrix} \,, \\[0.5em]
    \left<b\right>_k^{\text{High}} =& \begin{pmatrix}
        0.492 & \hspace{0.3cm} 0 & \hspace{0.3cm} 0.237 & \hspace{0.3cm} 0 & \hspace{0.3cm} - 0.254 & \hspace{0.3cm} 0
    \end{pmatrix} \,, \\[0.5em]
    \left<B\right>_{k\ell}^{\text{Low}} =&\begin{pmatrix}
        1.73 & \hspace{0.3cm} 0 & \hspace{0.3cm} 0.164 & \hspace{0.3cm} 0 & \hspace{0.3cm} 0 & \hspace{0.3cm} 0 \\[0.3em]
         & \hspace{0.3cm} 1.75 & \hspace{0.3cm} 0 & \hspace{0.3cm} 0.171 & \hspace{0.3cm} 0 & \hspace{0.3cm} 0 \\[0.3em]
          & \hspace{0.3cm}   & \hspace{0.3cm} 0.039 & \hspace{0.3cm} 0 & \hspace{0.3cm} 0 & \hspace{0.3cm} 0 \\[0.3em]
          & \hspace{0.3cm}   & \hspace{0.3cm}   & \hspace{0.3cm} 0.043 & \hspace{0.3cm} 0 & \hspace{0.3cm} 0 \\[0.3em]
          & \hspace{0.3cm}   & \hspace{0.3cm}   & \hspace{0.3cm}   & \hspace{0.3cm} 0.039 & \hspace{0.3cm} 0 \\[0.3em]
          & \hspace{0.3cm}   & \hspace{0.3cm}   & \hspace{0.3cm}   & \hspace{0.3cm}   & \hspace{0.3cm} 0.043
    \end{pmatrix} \, \text{ , and } \\[0.5em]
    \left<B\right>_{k\ell}^{\text{High}} =&\begin{pmatrix}
        0.134 & \hspace{0.3cm} 0 & \hspace{0.3cm} 0.063 & \hspace{0.3cm} 0 & \hspace{0.3cm} 0 & \hspace{0.3cm} 0 \\[0.3em]
         & \hspace{0.3cm} 0.145 & \hspace{0.3cm} 0 & \hspace{0.3cm} 0.069 & \hspace{0.3cm} 0 & \hspace{0.3cm} 0 \\[0.3em]
          & \hspace{0.3cm}   & \hspace{0.3cm} 0.030 & \hspace{0.3cm} 0 & \hspace{0.3cm} 0 & \hspace{0.3cm} 0 \\[0.3em]
          & \hspace{0.3cm}   & \hspace{0.3cm}   & \hspace{0.3cm} 0.033 & \hspace{0.3cm} 0 & \hspace{0.3cm} 0 \\[0.3em]
          & \hspace{0.3cm}   & \hspace{0.3cm}   & \hspace{0.3cm}   & \hspace{0.3cm} 0.030 & \hspace{0.3cm} 0 \\[0.3em]
          & \hspace{0.3cm}   & \hspace{0.3cm}   & \hspace{0.3cm}   & \hspace{0.3cm}   & \hspace{0.3cm} 0.033
    \end{pmatrix} \,.
\end{align}

\section{Constraints on Wilson Coefficients}\label{app:WCconst}
\begin{figure*}[h!]
    \centering
    \begin{subfigure}[t]{1\textwidth}
        \centering
        \includegraphics[width=1\textwidth]{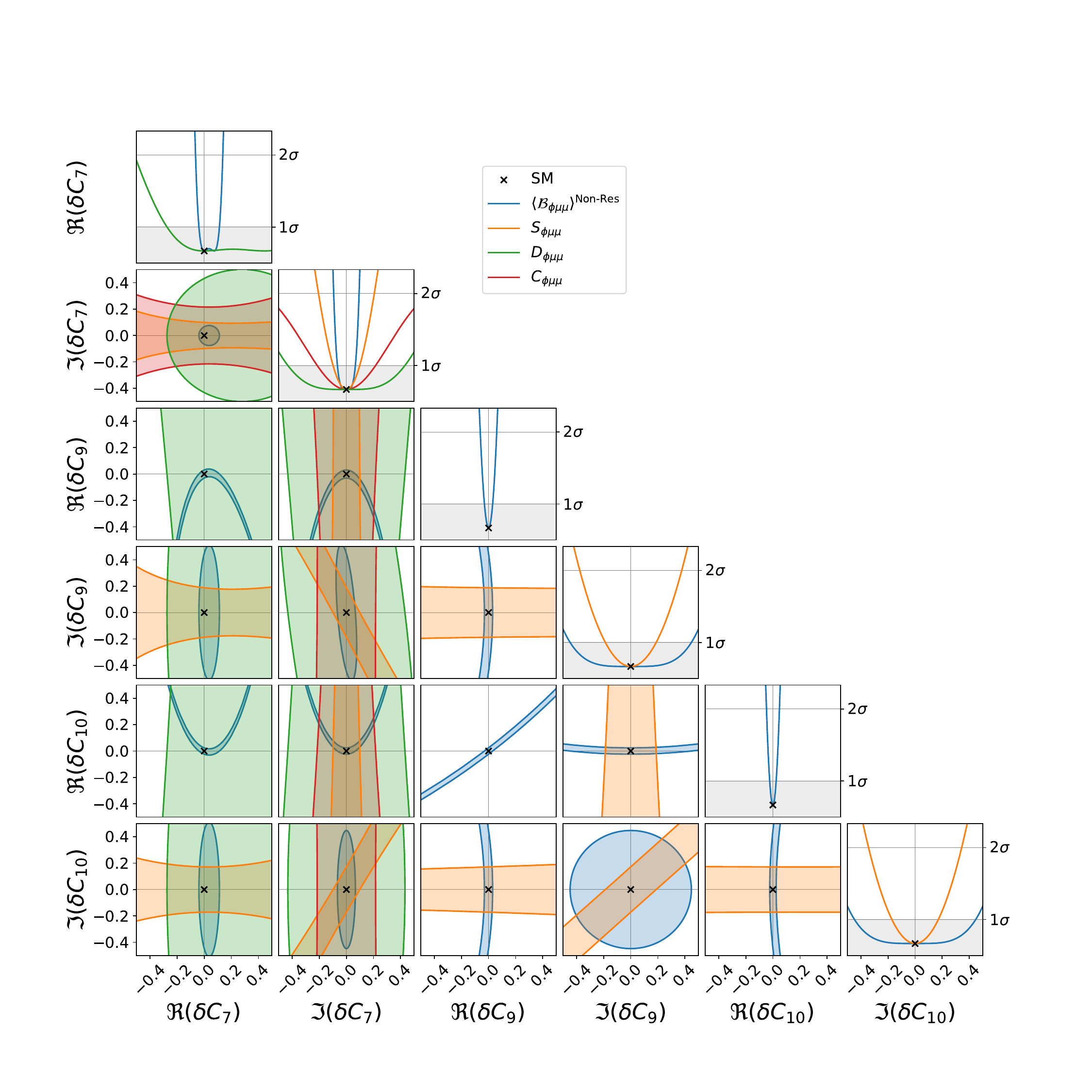}
    \end{subfigure}%
    \caption{The 2D constraints on the real and imaginary parts of the WCs are shown in the off-diagonal plots. The diagonal ones show the 1D likelihood of the corresponding WCs. The shaded regions represent $1\sigma$ contours, with different colors indicating constraints from various measurements.}
    \label{fig:WC_all}
\end{figure*}

\begin{figure*}[h!]
    \centering
    \begin{subfigure}[t]{1\textwidth}
        \centering
        \includegraphics[width=1\textwidth]{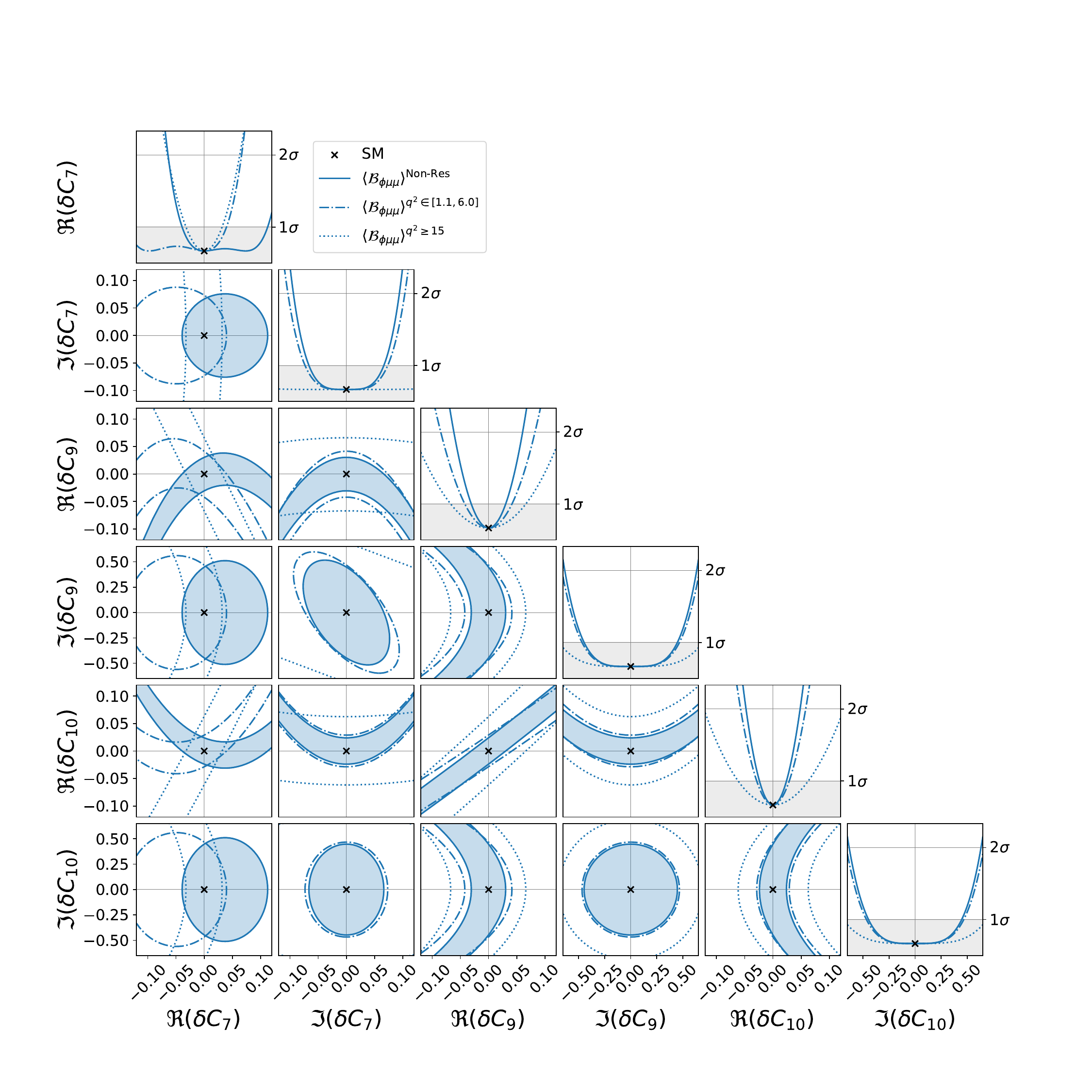}
    \end{subfigure}%
    \caption{The 2D constraints and the 1D likelihood of the corresponding WCs from Full, Low- and High-$q^2$ branching ratio measurements. The shaded regions represent $1\sigma$ contours of the Full $q^2$ measurement, and dot-dashed and dotted represent the Low- and High-$q^2$ measurements, respectively. The average of the three bins is shown in black solid line.}
    \label{fig:WC_q2}
\end{figure*}

The constraints on the WCs from additional observables are presented in Figures~\ref{fig:WC_all} and~\ref{fig:WC_q2}. Figure~\ref{fig:WC_all} displays the constraints from the branching ratio and all $C\!P$-violating observables: $S_{\phi\mu\mu}$, $D_{\phi\mu\mu}$, and $C_{\phi\mu\mu}$. Among these, $D_{\phi\mu\mu}$ and $C_{\phi\mu\mu}$ exhibit relatively lower sensitivity compared to the branching ratio and $S_{\phi\mu\mu}$. Figure~\ref{fig:WC_q2} illustrates the constraints obtained from different $q^2$ binning of the branching ratio measurement, highlighting that $\Re{(\delta C_7)}$ is particularly sensitive to the choice of $q^2$ binning. 
Theory errors are neglected in Figures~\ref{fig:WC_all} and~\ref{fig:WC_q2}.

\clearpage
\bibliographystyle{jhep}
\bibliography{biblio}
\end{document}